\documentclass[aps,prx,12pt,letterpaper,amsmath,amssymb]{revtex4-1}

\usepackage[utf8]{inputenc}
\usepackage{graphicx}
\usepackage{subfigure}
\usepackage[usenames,dvipsnames]{xcolor}
\usepackage{epstopdf}
\usepackage{amsmath,amsthm,amssymb}
\usepackage{latexsym}
\usepackage{bm}
\usepackage{dcolumn}
\usepackage{braket}
\usepackage[normalem]{ulem}
\usepackage{times}
\usepackage{hyperref}
\usepackage{enumitem}

\newcommand{\ignore}[1]{}
\newcommand{\boldfont}[1]{\textbf{#1}}
\newcommand{\figtextbf}[1]{#1}
\newcommand{\red}[1]{\textcolor{black}{#1}}

\newcommand{\boxedpph}{\boxed{++}}
\newcommand{\boxedmmh}{\boxed{--}}
\newcommand{\boxedpmh}{\boxed{+-}}
\newcommand{\boxedmph}{\boxed{-+}}

\begin{document}
\bibliographystyle{naturemag}

\title{Paradigm for approaching the forbidden phase transition in the one-dimensional Ising model at fixed finite temperature: Single chain in a magnetic field}
\author{Weiguo Yin}
\email{wyin@bnl.gov}
\affiliation{Condensed Matter Physics and Materials Science Division,
Brookhaven National Laboratory, Upton, New York 11973, USA}

\date{\today}

\begin{abstract}
In a previous paper [Weiguo Yin, \href{https://doi.org/10.1103/PhysRevResearch.6.013331}{Phys. Rev. Res. \red{\textbf{6}, 013331} (2024)}], the forbidden spontaneous phase transition in the one-dimensional Ising model was found to be approachable arbitrarily closely in decorated ladders by ultra-narrow phase crossover (UNPC) at a given finite temperature $T_0$ with the crossover width $2\delta T$ reduced exponentially, which resemble a genuine first-order transition with large latent heat. 
Here, I 
reveal that the forbidden phase transition can be approached at fixed $T_0$ as well in decorated single-chain Ising models in the presence of a magnetic field, in which $T_0$ is determined by the interactions involving only the decorated parts and the magnetic field, while $2\delta T$ is independently, exponentially reduced ($\delta T=0$ means a genuine transition) by restoring the ferromagnetic interaction between the ordinary spins on the chain backbone---which was neglected in the previous studies of pseudo-transition---thus manifesting that this asymptoticity to the forbidden transition is essentially the buildup of coherence in preformed crossover of local states. 
Furthermore, I show that the UNPC can be realized even in the absence of the conventional geometric frustration because the magnetic field itself can induce previously unnoticed hidden spin frustration. These findings make the doors wide open to the engineering and utilization of UNPC as a new paradigm for exploring exotic phenomena and 1D device applications. 


\end{abstract}


\maketitle

\section{Introduction}

The textbook Ising model 
describes collective behaviors such as phase transitions and critical phenomena in various physical, biological, economical, and social systems~\cite{Mattis_book_08_SMMS,Mattis_book_1985,Baxter_book_Ising,003_Huang_08_book}. Since Ernst Ising's proof one century ago~\cite{Ising1925}, it has been well-known that phase transition at finite temperature does not exist in the Ising model with short-range interactions in one dimension~\cite{Cuesta_1D_PT}. 
Yet, little is known about whether this forbidden transition could be approached arbitrarily closely---at fixed finite temperature $T_0$---until recently such asymptoticity was successfully found in decorated ladder Ising models in the absence of a magnetic field~\cite{Yin_MPT,Yin_icecreamcone}. 
On the other hand, in the presence of a magnetic field, ultra-narrow phase crossover termed as ``pseudo-transition'' was found in decorated single-chain Ising models with strong geometric frustration~\cite{005_Galisova_PRE_15_double-tetrahedral-chain,
007_Torrico_PRA_16_Ising-XYZ-diamond-chain,
009_review_Souza_SSC_18_Ising-XYZ-diamond-chain_double-tetrahedral-chain-spin-electron,
010_Carvalho_JMMM_18_Ising-XYZ-diamond-chain_quantum-entanglement,
011_Rojas_BJP_20_Ising-Heisenberg-tetrahedral_diamond,
013_Rojas_PRE_19_previous_4_models,
014_Rojas_JPC_20_Ising–Heisenberg_spin-1-double-tetrahedral-chain,
015_Strecka_APPA_20_Ising-diamond-chain,
015-7_Canova_CzechoslovakJP_04_Ising–Heisenberg_diamond_chain,
015-8_Canova_JPC_06_Ising–Heisenberg-spin-S-diamond-chain,
017_Krokhmalskii_PA_21_3-previous-chains_effective_model,
016_Strecka_book_chapter}; however, there is no hint to the question of how to make the crossover width $2\delta T$ narrower and narrower while keeping $T_0$ unchanged, since $2\delta T$ was not even defined in terms of the model parameters. The independent control of $T_0$ and $2\delta T$ by different interactions will not only help push the limit in our understanding of phase transitions arbitrarily close to the forbidden regime, but also provide promising potentials in technology applications. It is thus imperative to explore whether and how the forbidden in-field phase transition could be approached arbitrarily closely at a given $T_0$ in the single-chain Ising models. 

The one-dimensional (1D) Ising model on a decorated single chain is generally defined as $H=H_\mathrm{ordinary}+\sum_{i} H^{(i)}_\mathrm{decorated}$, where
\begin{eqnarray}
H_\mathrm{ordinary}&=&-J\sum_{i=1}^{N}\sigma_{i}\sigma_{i+1}-h\mu_a \sum_{i=1}^{N}\sigma_{i} \label{ordinary} \end{eqnarray}
describes the ordinary single chain without the decoration (Fig.~1\boldfont{a}) with $\sigma_{i}=\pm 1$ standing for the $i$th \emph{ordinary} spin---in fact, it can be used to describe any two-value system, eg. open or close in neural networks~\cite{Schneidman2006}, yes or no in voting. $N$ is the total number of the ordinary spins and $\sigma_{N+1}\equiv\sigma_{1}$ (i.e., the periodic boundary condition). $J$ is the interaction between nearest-neighboring ordinary spins. $h$ depicts the magnetic field and $\mu_a$ the magnetic moment of the ordinary spins.  $H^{(i)}_\mathrm{deorated}$ describes the decorated part in between the $i$th and $(i+1)$th ordinary spins (Fig.~1\boldfont{b}), which can be any finite-size subsystem as long as it couples to the two nearest ordinary spins by the Ising-type interactions only. To date, the simplest $H^{(i)}_\mathrm{deorated}$ considered in the literature of ``pseudo-transition'' is the Ising diamond~\cite{016_Strecka_book_chapter} (Fig.~1\boldfont{d}) given by
\begin{eqnarray}
H^{(i)}_\mathrm{deorated}&=&-J_1(\sigma_{i}+\sigma_{i+1})\sum_{k=1,2} s_{i,k}-J_2 s_{i,1}s_{i,2}- h\mu_b\sum_{k=1,2} s_{i,k},
\label{diamond}
\end{eqnarray}
where $s_{i,k}=\pm1$ denotes the $k$th decorated Ising spin for the $i$th bond of the ordinary chain or the backbone. $\mu_b$ is the magnetic moment of the decorated spins. The antiferromagnetic $J_1<0$ and $J_2<0$ were considered; they form triangles yielding strong geometric frustration~\cite{Balents_nature_frustration,Miyashita_10_review_frustration} near $J_2/J_1=2$~\cite{016_Strecka_book_chapter}.

It is instrumental to start with a comparative review of (i) the previous studies of the decorated Ising chains with pseudo-transition in the presence of magnetic field~\cite{005_Galisova_PRE_15_double-tetrahedral-chain,
007_Torrico_PRA_16_Ising-XYZ-diamond-chain,
009_review_Souza_SSC_18_Ising-XYZ-diamond-chain_double-tetrahedral-chain-spin-electron,
010_Carvalho_JMMM_18_Ising-XYZ-diamond-chain_quantum-entanglement,
011_Rojas_BJP_20_Ising-Heisenberg-tetrahedral_diamond,
013_Rojas_PRE_19_previous_4_models,
014_Rojas_JPC_20_Ising–Heisenberg_spin-1-double-tetrahedral-chain,
015_Strecka_APPA_20_Ising-diamond-chain,
015-7_Canova_CzechoslovakJP_04_Ising–Heisenberg_diamond_chain,
015-8_Canova_JPC_06_Ising–Heisenberg-spin-S-diamond-chain,
017_Krokhmalskii_PA_21_3-previous-chains_effective_model,
016_Strecka_book_chapter} and (ii) the recent investigations of the decorated Ising ladders in the absence of the magnetic field~\cite{Yin_MPT,Yin_icecreamcone,Hutak_PLA_21_trimer}. In particular, we ask questions as to what the pseudo-transition research has done and has not done, compared with what we have learned from the spontaneous ultra-narrow phase crossover in the ladder. This task is greatly simplified by a recent summary~\cite{017_Krokhmalskii_PA_21_3-previous-chains_effective_model} of the pseudo-transition research in the effective Hamiltonian approach, where tracing out the decorated parts results in the ordinary Ising-chain model with temperature-dependent parameters $J_\mathrm{eff}(T)$ and $h_\mathrm{eff}(T)$ in place of $J$ and $h$ in Eq.~(\ref{ordinary}). Two key conclusions about the existence of the pseudo-transition were reached~\cite{017_Krokhmalskii_PA_21_3-previous-chains_effective_model}: (1) $h_\mathrm{eff}(T)$ must experience the sign change as a function of temperature $T$ and $T_0$ is determined by $h_\mathrm{eff}(T)=0$. (2) The decoration was done to create geometrical frustration so that the system's first low-lying excited state has much higher degeneracy and just slightly higher energy than the ground state, then an entropy-driven crossover between them would occur at finite temperature~\cite{005_Galisova_PRE_15_double-tetrahedral-chain,
007_Torrico_PRA_16_Ising-XYZ-diamond-chain,
009_review_Souza_SSC_18_Ising-XYZ-diamond-chain_double-tetrahedral-chain-spin-electron,
010_Carvalho_JMMM_18_Ising-XYZ-diamond-chain_quantum-entanglement,
011_Rojas_BJP_20_Ising-Heisenberg-tetrahedral_diamond,
013_Rojas_PRE_19_previous_4_models,
014_Rojas_JPC_20_Ising–Heisenberg_spin-1-double-tetrahedral-chain,
015_Strecka_APPA_20_Ising-diamond-chain,
015-7_Canova_CzechoslovakJP_04_Ising–Heisenberg_diamond_chain,
015-8_Canova_JPC_06_Ising–Heisenberg-spin-S-diamond-chain,
017_Krokhmalskii_PA_21_3-previous-chains_effective_model,
016_Strecka_book_chapter}. This physics of phase crossover is rather generic~\cite{Miyashita_10_review_frustration}. To make the crossover ultra-narrow, it is of normal practice to place the system close to the critical point or the phase boundary in the zero-temperature phase diagram. It was found that $T_0 \to 0$ as the pseudo-transition is ``tracked down in the critical point of the standard Ising-chain model at $h = 0$ and $T = 0$''~\cite{017_Krokhmalskii_PA_21_3-previous-chains_effective_model}. In other words, 
\begin{equation}
T_0  \to 0 \;\; \mathrm{as} \;\; 2\delta T \to 0,
\label{T0-dT}
\end{equation}
in this traditional paradigm of realizing the ultra-narrow phase crossover by approaching the zero-temperature phase boundary of two competing phases. Therefore, the pseudo-transition does not appear to support our goal of approaching the forbidden phase transition at fixed finite $T_0$. 
Now that the knowledge of how to realizing the goal in the Ising ladders has become available~\cite{Yin_MPT,Yin_icecreamcone}, we understand that the following key pieces of information were missing in the previous studies of pseudo-transition \emph{and} we are able to quickly find the solution in the term of ``coherent pseudo-transition'' (CPT) that unifies the ultra-narrow phase crossovers in both the ladder and the single-chain Ising models: 

1) \emph{On the description of the targeted phenomenon.---}The crossover width $2\delta T$ was never clearly defined and expressed by the model parameters (note that $\delta T=0$ means a genuine phase transition) for the pseudo-transition, while it was always presented in terms of specific heat, entropy, magnetic susceptibility, or overall magnetization~\cite{005_Galisova_PRE_15_double-tetrahedral-chain,
007_Torrico_PRA_16_Ising-XYZ-diamond-chain,
009_review_Souza_SSC_18_Ising-XYZ-diamond-chain_double-tetrahedral-chain-spin-electron,
010_Carvalho_JMMM_18_Ising-XYZ-diamond-chain_quantum-entanglement,
011_Rojas_BJP_20_Ising-Heisenberg-tetrahedral_diamond,
013_Rojas_PRE_19_previous_4_models,
014_Rojas_JPC_20_Ising–Heisenberg_spin-1-double-tetrahedral-chain,
015_Strecka_APPA_20_Ising-diamond-chain,
015-7_Canova_CzechoslovakJP_04_Ising–Heisenberg_diamond_chain,
015-8_Canova_JPC_06_Ising–Heisenberg-spin-S-diamond-chain,
017_Krokhmalskii_PA_21_3-previous-chains_effective_model,
016_Strecka_book_chapter}; these physical quantities are derivatives of the free energy with respect to the global parameter $T$ or $h$ and thus depend on the details of the model. For spontaneous CPT in decorated Ising ladders, the on-rung parent-spin correlation function was identified as the order parameter (OP) that has a well-defined value space of $[-1,1]$ with the value $0$ meaning $T_0$ and its inverse slope at $T_0$ meaning $\delta T$, characterizing the CPT as an abrupt change in the OP between nearly $-1$ to nearly $+1$~\cite{Yin_MPT,Yin_icecreamcone}. Moreover, the general form of the OP does not depend on the details of the model; its mathematical derivation and numerical computation can be easily carried out. Such an OP provides an accurate, convenient, and microscopic description of the CPT; its identification greatly accelerated the search for CPT. Here for the decorated single-chain Ising models, the OP that has the same features is $\langle \sigma_i \rangle$, the magnetization of the ordinary spins on the chain backbone (not the overall magnetization that includes the decorated parts): Its sign change and zero value at $T_0$ is consistent with the behavior of $h_\mathrm{eff}(T)$; its inverse derivative at $T_0$ defines $\delta T = \left|\partial \langle \sigma_i \rangle/\partial T\right|^{-1}_{T=T_0}$ (Fig.~\ref{Fig:OP}a).

2) \emph{On the model Hamiltonian.---}Surprisingly, the $J$ term---the Ising interaction between the ordinary spins on the chain backbone (red bonds in Figs.~1\boldfont{a}, 1\boldfont{b}, 1\boldfont{e}, 1\boldfont{f})---was neglected in the previous studies of pseudo-transition~\cite{005_Galisova_PRE_15_double-tetrahedral-chain,
007_Torrico_PRA_16_Ising-XYZ-diamond-chain,
009_review_Souza_SSC_18_Ising-XYZ-diamond-chain_double-tetrahedral-chain-spin-electron,
010_Carvalho_JMMM_18_Ising-XYZ-diamond-chain_quantum-entanglement,
011_Rojas_BJP_20_Ising-Heisenberg-tetrahedral_diamond,
013_Rojas_PRE_19_previous_4_models,
014_Rojas_JPC_20_Ising–Heisenberg_spin-1-double-tetrahedral-chain,
015_Strecka_APPA_20_Ising-diamond-chain,
015-7_Canova_CzechoslovakJP_04_Ising–Heisenberg_diamond_chain,
015-8_Canova_JPC_06_Ising–Heisenberg-spin-S-diamond-chain,
017_Krokhmalskii_PA_21_3-previous-chains_effective_model,
016_Strecka_book_chapter}, that is, Fig.~1\boldfont{c} was studied instead of the more general Fig.~1\boldfont{b}. A possible reason for the omission of $J$ could be that the standard geometric frustration from the triangles formed by the antiferromagnetic bonds is more obvious for $J=0$, as shown in Fig.~1d for the Ising diamond chain, the hitherto simplest model with pseudo-transition. However, the CPT for the Ising ladders tells us that the on-leg decoration (which controls $\delta T$) can be done independently of the on-rung decoration (which controls $T_0$)~\cite{Yin_MPT,Yin_icecreamcone}. We shall use the $J\ne 0$ Ising diamond chain model (Fig.~1\boldfont{e}) to show that similarly, the $J$ term independently exponentially reduces the crossover width $2\delta T$ for fixed finite $T_0$, since $J$ has no effect on $h_\mathrm{eff}(T)$ but is a separate addend in $J_\mathrm{eff}(T)$, i.e.,  $J_\mathrm{eff}(T,J) = J + J_\mathrm{eff}(T,0)$ (see Eq.~(\ref{effective}) in the \textbf{Method}).

3) \emph{On the underlying mechanism.---}Only the frustration of geometric frustration type (namely the lattice has frustration regardless of the presence or absence of the magnetic field) was considered for the pseudo-transition. \ignore{Using the effective Hamiltonian approach to study the same spontaneous CPT in the decorated Ising ladders with geometrically frustrated triangles as Refs.~\cite{Yin_MPT,Yin_icecreamcone}, Hutak \textsl{et al.} conjectured that geometric frustration is not necessarily a prerequisite for the realization of spontaneous CPT in such ladders~\cite{Hutak_PLA_21_trimer}. Although this conjecture has not been proved for the ladder yet, 
we were inspired to extend the conjecture to the decorated single-chain Ising models, because} It was recently emphasized~\cite{Yin_g} that the magnetic field on its own could induce \emph{hidden} spin frustration in ferrimagnet-like systems without geometric frustration~\cite{Bell_JPC_74_Ising_ferri_1}. \ignore{It suffices to prove the conjecture with one example of the systems without geometric frustration;} Here we show the existence of CPT in the Ising diamond chain for $J_2=0$ (Fig.~1\boldfont{f}), where the triangles formed by the $J$ and $J_1$ bonds are not frustrated because $J>0$ is ferromagnetic. The hidden high degeneracy generated by the magnetic field, not by geometric frustration, is explained in the ground-state phase diagram of the Ising diamond chain (the red circle on the red dashed line in Fig.~\ref{Fig:zeroT}). \red{An immediate improvement is the dramatic increase in $T_0$, e.g., by 2200\% when $J_2/J_1$ is moved away from $1.95$ to $0$ (note that $J_2/J_1=2$ sets the phase boundary), as shown in Fig.~\ref{Fig:T-J2}(b) and Figs.~\ref{Fig:T-h}(b)(d).}

These findings thoroughly expose the mathematical structure of CPT as a generic mechanism for generating ultra-narrow phase crossover at fixed $T_0$ in the 1D Ising models. One can use various decorations to yield even very broad phase crossovers and then use interactions that enhance the coherence of the order parameter to turn the broad phase crossover to be the CPT---that is how CPT (coherent pseudo-transition) got its name---reminiscent of the notion of preformed pairs and their coherence buildup in the field of high-temperature superconductivity~\cite{Emery_Nature_95}. Given the physical effects that the CPT resembles a genuine first-order phase transition with large latent heat and the fact that 
 the Ising model has already been implemented in electronic circuits~\cite{Ising_FPGA}, optical 
networks~\cite{Pierangeli_IsingMachine_PRL19},  
and optical lattices~\cite{Bernien_Nature_17_Rydberg},   
the CPT-based 1D devices for thermal applications appear to be feasible. The features that $T_0$ and $2\delta T$ can be independently controlled by different parameters and different decoration methods could be attractive in engineering 1D thermal sensors, for example. The doors to the engineering and utilization of CPT are now wide open. 

\section{RESULTS AND DISCUSSIONS}

We describe the mathematical details in the Method section and show key results below. The OP is given by
\begin{equation}
    \langle \sigma_i \rangle = \frac{\sinh(\beta h_\mathrm{eff}\mu_a)}{\sqrt{\sinh^2(\beta h_\mathrm{eff}\mu_a)+e^{-4\beta J_\mathrm{eff}}}}. \label{OP_eff_main_text}
\end{equation}
Clearly, $\langle \sigma_i \rangle \in [-1,1]$. $\langle \sigma_i \rangle=0$ when $h_\mathrm{eff}=0$ at $T_0$ and for fixed $T_0$, $\delta T \propto e^{-2 J_\mathrm{eff}/k_\mathrm{B}T_0}$ exponentially decays as $J$, an addend in $J_\mathrm{eff}$, increases. 

The $T-J_2$ phase diagrams from the density plot of $\langle\sigma_i\rangle$ for the previously studied model (i.e., $J_2/J_1=1.95$ and $J=0$) ~\cite{015_Strecka_APPA_20_Ising-diamond-chain,016_Strecka_book_chapter} is presented in Fig.~\ref{Fig:T-J2}a. \red{For $J=0$,} the sharp phase crossover from $\langle\sigma_i\rangle=-1$ to $+1$ occurs only near the small region with strong geometric frustration ($J_2/J_1=2$ is the FRI-FRU phase boundary; cf. Fig.~\ref{Fig:zeroT}). The resultant $T_0$ is very low, e.g.,  $k_\mathrm{B}T_0S^2/|J_1|\approx 0.036$ for $J_2/J_1=1.95$ \red{[c.f., Equation~(\ref{T0-dT})]}. As $J_2/J_1$ decreases, $T_0$ increases ($k_\mathrm{B}T_0S^2/|J_1|\approx 0.821$ for $J_2=0$) while the density is spread over a wider temperature interval, meaning larger $\delta T$. However, for the present new model with $J> 0$, Fig.~\ref{Fig:T-J2}b shows that all the broad crossovers have been turned to be ultranarrow by increasing $J$. The same is observed in the $T-h$ phase diagrams presented in Figs.~\ref{Fig:T-h}a and \ref{Fig:T-h}b for $J_2/J_1=1.95$ and Figs.~\ref{Fig:T-h}c and \ref{Fig:T-h}d for $J_2/J_1=0$.

To gain more insights, we present the $T$ dependence of $h_\textrm{eff}$ and $J_\textrm{eff}$ in Fig.~\ref{Fig:OP}b and \ref{Fig:OP}d, respectively, for $J=0$ and several $J_2/J_1$, as well as $T_0$ and $\delta T$ as a function of $J_2/J_1$ in Fig.~\ref{Fig:OP}c. When $J=0$,  $J_\textrm{eff}/k_\mathrm{B}T_0\approx 6.58$ for $J_2/J_1=1.95$ resulting in $\delta T\approx 2\times 10^{-7}$, but the value quickly drops to $J_\textrm{eff}/k_\mathrm{B}T_0 \approx0.13$ for $J_2=0$ resulting in $\delta T \approx 1.53$.
Nevertheless, as we turn on $J$, $\delta T$ can be made narrower and narrower not only for the $J_2/J_1=1.95$ already ultra-narrow case (Fig.~\ref{Fig:OP}e) but also for the initially wide crossover for $J_2=0$ (Fig.~\ref{Fig:OP}f) with $T_0$ fixed in both cases. However, the underlying microscopic mechanisms for the former case with strong geometric frustration and the latter case without geometric frustration are quite different, as elaborated below.

As shown in Fig.~\ref{Fig:thermal}, both cases resemble a genuine first-order phase transition with the entropy jump and gigantic susceptibility at $T_0$, but the entropy per unit cell is flattened at $\ln 2$ and $2\ln 2$ for $J_2/J_1=1.95$ (Fig.~\ref{Fig:thermal}a) and $J_2=0$ (Fig.~\ref{Fig:thermal}b), respectively.  The former is expected since the case with $J_2/J_1=1.95$ is located near the FRI-FRU phase boundary in the ground-state phase diagram (the cyan circle in Fig.~\ref{Fig:zeroT}). The FRU phase has the degeneracy of two per unit cell due to the frustrated decorated spins; the crossover is driven by this entropy gain by $\ln2$ per unit cell~\cite{016_Strecka_book_chapter}, while the ordinary spins flip from $\sigma_i\approx-1$ to $+1$ still being locked together by large $J_\mathrm{eff}/k_\mathrm{B}T_0$. By sharp contrast, the case with $J_2=0$ is far away from any phase boundary and the entropy flattening at $2\ln 2$ is astonishing. This exotic phenomenon originates from a \emph{hidden} remarkably high degeneracy induced by the magnetic field even in the system without geometric frustration: As shown by the red circle in Fig.~\ref{Fig:zeroT}, the case with $J_2=0$ is seated right on the extended line of the FRU-SPP phase boundary (red dashed line). This means that once the system is heated out of the FRI ground state, it will be frustrated in choosing between FRU or SPP, resulting in the effective decoupling of the two decorated spins per unit cell from the lattice---that is the entropy gain of $2\ln2$. Meanwhile, the ordinary spins flip from $\sigma_i\approx-1$ to $+1$ and are still locked together by large $J_\mathrm{eff}/k_\mathrm{B}T_0$. In other words, it is this lockup, ``calm,'' or the buildup of coherence in the order parameter $\sigma_i$ of the ordinary spins that makes the decorated spins fully frustrated in response to the heated atmosphere. This is a distinctly new mechanism for driving the CPT. It is also opposite to the zero-temperature critical point recently emphasized as the ``half-ice half-fire'' state in ferrimagnet-like systems where the ordinary spins are fully frustrated while the decorated spins are forced to be calm by the critical magnetic field~\cite{Yin_g}. Now the FRI regime of the ground-state phase diagram has two paths toward CPT through either the geometric-frustration-driven FRI-FRU or the hidden-frustration-driven FRU-SPP phase boundary. As demonstrated in the density plot of the entropy right above $T_0$ (Fig.~\ref{Fig:h-J2}a) and that of the entropy jump at $T_0$ (Fig.~\ref{Fig:h-J2}b), the entropy jump of about $\ln 2$ to $2\ln 2$ takes place in most areas of the FRI regime except for weak $h$ and the largest jump occurs approximately along the hidden frustration line. This difference in the entropy jump together with $T_0$ and $\delta T$ could be used to train deep neural networks to predict the model parameters of the decorated 1D Ising model with CPT~\cite{Yin_PRB_23_ML_t-J}.

\section{Summary}

In summary, a simple and general way by including the ferromagnetic $J$ term is found to not only transform all the previously studied systems---with geometric frustration---in the context of pseudo-transition into the \red{ultranarrow phase crossover that possesses} the highly desirable features: $T_0$ and $2\delta T$ can be independently controlled by different parameters and different decoration methods, but also unexpectedly expose the hidden field-induced frustration to generate the \red{ultranarrow phase crossover} in decorated Ising chains without the conventional geometric frustration. With the discoveries of both the spontaneous \red{ultranarrow phase crossover} in the decorated Ising two-leg ladders (which is DNA-like) and the field-driven \red{ultranarrow phase crossover} in the decorated Ising single chains (which is RNA-like), the foundation of the research direction \red{in ultranarrow phase crossover} has been solidly established. Given the prominent roles of the Ising model and frustration in understanding collective phenomena in various physical, biological, economical, and social systems, and the prominent roles of 1D systems in research, education, and technology applications, as well as the recent technological advancement that the Ising model has already been implemented in various physical systems~\cite{Ising_FPGA,Pierangeli_IsingMachine_PRL19,Bernien_Nature_17_Rydberg},    
we anticipate that the present new insights to phase transitions and frustration effects will stimulate further research and development about \red{ultranarrow phase crossover} and its applications. 


\begin{acknowledgments}
The author is grateful to D. C. Mattis for mailing him a copy of Ref.~\cite{Mattis_book_08_SMMS} as a gift and inspiring discussions over the years. 
Brookhaven National Laboratory was supported by U.S. Department of Energy (DOE) Office of Basic Energy Sciences (BES) Division of Materials Sciences and Engineering under contract No. DE-SC0012704.
\end{acknowledgments}

\appendix
\section{The Method}
\label{method}
The central quantity of statistic mechanics is  the partition function $Z=\mathrm{Tr}\left( e^{-\beta H} \right)$ where $H$ is the Hamiltonian of the system and $\beta=1/(k_\mathrm{B}T)$ with $T$ being the temperature and 
$k_\mathrm{B}$ the Boltzmann constant~\cite{Mattis_book_08_SMMS}. The free energy \red{per unit cell} $f(T)=-\red{\frac{1}{N}}k_\mathrm{B}T\ln Z$ \red{(where $N$ is the number of unit cells)} determines many important thermodynamic properties such as the entropy $S=-\partial f/\partial T$, the specific heat $C_v=T\partial S/\partial T$, the magnetization $m=-\partial f/\partial h$, and the magnetic susceptibility $\chi=\partial m/\partial h$. Here the OP is $\langle \sigma_i \rangle = -(1/h)\partial f/\partial\mu_a$.

The partition function of a 1D Ising model can be obtained exactly by using the transfer matrix method~\cite{Mattis_book_08_SMMS,Mattis_book_1985,Baxter_book_Ising,003_Huang_08_book,005_Galisova_PRE_15_double-tetrahedral-chain,
007_Torrico_PRA_16_Ising-XYZ-diamond-chain,
009_review_Souza_SSC_18_Ising-XYZ-diamond-chain_double-tetrahedral-chain-spin-electron,
010_Carvalho_JMMM_18_Ising-XYZ-diamond-chain_quantum-entanglement,
011_Rojas_BJP_20_Ising-Heisenberg-tetrahedral_diamond,
013_Rojas_PRE_19_previous_4_models,
014_Rojas_JPC_20_Ising–Heisenberg_spin-1-double-tetrahedral-chain,
015_Strecka_APPA_20_Ising-diamond-chain,
015-7_Canova_CzechoslovakJP_04_Ising–Heisenberg_diamond_chain,
015-8_Canova_JPC_06_Ising–Heisenberg-spin-S-diamond-chain,
017_Krokhmalskii_PA_21_3-previous-chains_effective_model,
016_Strecka_book_chapter,Yin_MPT,Yin_icecreamcone,Yin_g,Bell_JPC_74_Ising_ferri_1} 
and is given by 
\begin{equation}
Z=\mathrm{Tr}\left(\Lambda^N\right)=\sum_k{\lambda_k^N} \;\;\to\;\; \lambda^N \;\mathrm{for}\; N\to \infty,
\end{equation} 
where $\Lambda$ is the transfer matrix, $\lambda_k$ the $k$th eigenvalue of $\Lambda$, and $\lambda$ the largest eigenvalue. Thus, in the thermodynamic limit, the free energy per unit cell $f(T)=-\lim^{}_{N\to\infty}\frac{1}{N} k_\mathrm{B}T\ln Z=-k_\mathrm{B}T\ln\lambda$. 

To calculate the partition function $Z=\mathrm{Tr}\left( e^{-\beta H} \right)$ for the general model  of the decorated Ising chains defined in Eq.~(\ref{ordinary}) and illustrated in Fig.~1\boldfont{b}, the decorated sites can be exactly summed out as they are coupled only to the nearest neighboring ordinary spins, yielding \emph{the decoration's  contribution functions}
\begin{equation}
\boxed{\pm\pm}_{i}=\sum 
\left(e^{\beta H^{(i)}_\mathrm{decorated}}\right)_{{\sigma_{i}=\pm 1},\;\sigma_{i+1}=\pm 1}
\label{rainbow}
\end{equation}
where the sum is over all possible states made up by the decorated subsystem for one of the four combinations of $(\sigma_{i}$, $\sigma_{i+1})=(\pm 1, \pm 1)$.
They are translationally invariant, i.e., $\boxed{\pm\pm}_{i}=\boxed{\pm\pm}$. 

The transfer matrix is of the following form:
\begin{eqnarray}
\Lambda_\mathrm{single-chain}=\left(
\begin{array}{cc}
  a & c \\
  c & b \\
\end{array}
\right)=\left(
\begin{array}{cc}
  e^{\beta J+\beta h \mu_a} \boxedpph & e^{-\beta J} \boxedpmh \\
  e^{-\beta J} \boxedmph & e^{\beta J-\beta h \mu_a} \boxedmmh \\
\end{array}
\right),
\label{single}
\end{eqnarray}
Its largest eigenvalue is
\begin{equation}
\lambda=\frac{1}{2}e^{\beta J}\left[\Upsilon_+ + \sqrt{\Upsilon_-^2+4e^{-4\beta J}\boxedpmh^2}\right],
\label{lamda}
\end{equation}
with \emph{the frustration functions} 
\begin{eqnarray}
\Upsilon_\pm&=&e^{\beta \mu_a h}\boxedpph \pm e^{-\beta \mu_a h}\boxedmmh,
\label{Y}
\end{eqnarray}
which is independent of $J$.

The crossing of $a$ and $b$ in Eq.~(\ref{single}) occurs when $\Upsilon_-$ changes sign at $T_0$ where $\Upsilon_-=0$.
\textbf{This means that $T_0$ is independent of $J$.} 
Meanwhile, the $\boxedpmh$ term in Eq.~(\ref{lamda}) has a prefactor of $e^{-4\beta J}$, \textbf{which exponentially decreases to zero as ferromagnetic $J>0$ increases for fixed finite $T_0$}; thus, if Eq.~(\ref{lamda}) is approximated by neglecting the $\boxedpmh$ term inside $\sqrt{\cdots}$,
\begin{equation}
\lambda \simeq \frac{1}{2}e^{\beta J}(\Upsilon_+ + |\Upsilon_-|),
\label{lamda2}
\end{equation}
which becomes non-analytic. The difference between Eq.~(\ref{lamda}) and Eq.~(\ref{lamda2}) takes place in a region of $(T_0-\delta T, T_0 +\delta T)$, where the crossover width $2\delta T$ can be estimated by $|\Upsilon_-|=2e^{-2\beta J}\boxedpmh$ at $T_0\pm\delta T$. Following Ref.~\red{\onlinecite{Yin_MPT}}, an alternative and consistent way to measure $\delta T$ is to find such an order parameter that has a well-defined value space of $[-1,1]$ with the value $0$ meaning $T_0$ and its inverse slope at $T_0$ meaning $\delta T$. It is the magnetization of the ordinary spins given by
\begin{equation}
    \langle \sigma_i \rangle = -\frac{\partial f}{h\,\partial \mu_a} = \frac{\Upsilon_-}{\sqrt{\Upsilon_-^2+4e^{-4\beta J}\boxedpmh^2}}. \label{OP_general}
\end{equation}
\begin{equation}
    \delta T = \left(\frac{\partial \langle \sigma_i \rangle}{\partial T}\right)^{-1}_{T=T_0} = e^{-\frac{2J}{k_\mathrm{B}T_0}}\frac{4k_\mathrm{B}T_0^2 \boxedpmh}{-\Upsilon_+}\left[2h\mu_a+\left(\frac{\partial \ln \boxedpph}{\partial \beta}-\frac{\partial \ln \boxedmmh}{\partial \beta}\right)\right]^{-1}_{\beta=\frac{1}{k_\mathrm{B}T_0}}. 
    \label{dT_general}
\end{equation}
Again, it is clear that the crossover width $2\delta T$ decreases exponentially as $J$ increases for fixed finite $T_0$. This order parameter provides an accurate, convenient, and microscopic description of CPT. The use of the CPT order parameter accelerates the finding of CPT. 

The above exact solution can also be represented in terms of temperature-dependent effective interactions and field on the ordinary spins~\cite{017_Krokhmalskii_PA_21_3-previous-chains_effective_model,Yin_MPT,Yin_icecreamcone,Hutak_PLA_21_trimer}:
\begin{eqnarray}
h_\mathrm{eff}&=&\frac{1}{2\beta \mu_a}(\ln a- \ln b)= h + \frac{1}{2\beta\mu_a}\left(\ln \boxedpph- \ln \boxedmmh\right), \label{effective}\\
J_\mathrm{eff}&=&\frac{1}{2\beta}\left[\frac{1}{2}(\ln a +\ln b)-\ln c\right]=J+\frac{1}{2\beta}\left[\frac{1}{2}\left(\ln \boxedpph+ \ln \boxedmmh\right)-\ln \boxedpmh\right],\nonumber \\
A&=&\frac{1}{2\beta}\left[\frac{1}{2}(\ln a +\ln b)+\ln c\right]=\frac{1}{2\beta}\left[\frac{1}{2}\left(\ln \boxedpph+ \ln \boxedmmh\right)+\ln \boxedpmh\right].\nonumber 
\end{eqnarray} 
Note that $\mu_a$ appears in $h_\mathrm{eff}$ only and $J$ appears in $J_\mathrm{eff}$ only---with $J_\mathrm{eff}(T,J) = J + J_\mathrm{eff}(T,0)$. This means that $J$ has no impact on the determination of $T_0$, therefore it can be used to change $\delta T$ for fixed $T_0$.
The resulting transfer matrix is expressed by
\begin{eqnarray}
\Lambda_\mathrm{single-chain}=\left(
\begin{array}{cc}
  a & c \\
  c & b \\
\end{array}
\right)=e^{\beta A}\left(
\begin{array}{cc}
  e^{\beta J_\mathrm{eff}+\beta h_\mathrm{eff} \mu_a} & e^{-\beta J_\mathrm{eff}} \\
  e^{-\beta J_\mathrm{eff}} & e^{\beta J_\mathrm{eff}-\beta h_\mathrm{eff} \mu_a} \\
\end{array}
\right),
\label{single_eff}
\end{eqnarray}
Its largest eigenvalue is
\begin{equation}
\lambda=e^{\beta A}e^{\beta J_\mathrm{eff}}\left[\cosh(\beta h_\mathrm{eff}\mu_a) + \sqrt{\sinh^2(\beta h_\mathrm{eff}\mu_a)+e^{-4\beta J_\mathrm{eff}}}\right].
\label{lamda_eff}
\end{equation}
$T_0$ is determined by $\sinh(\beta h_\mathrm{eff}\mu_a)=0$, i.e., $h_\mathrm{eff}(T_0)=0$~\cite{017_Krokhmalskii_PA_21_3-previous-chains_effective_model}, which is the same as  $\Upsilon_-(T_0)=0$. The order parameter is
\begin{equation}
    \langle \sigma_i \rangle = -\frac{\partial f}{\partial (h\mu_a)} =-\frac{\partial f}{\partial (h_\mathrm{eff}\mu_a )}= \frac{\sinh(\beta h_\mathrm{eff}\mu_a)}{\sqrt{\sinh^2(\beta h_\mathrm{eff}\mu_a)+e^{-4\beta J_\mathrm{eff}}}}. \label{OP_eff}
\end{equation}
\begin{equation}
    \delta T = \left(\frac{\partial \langle \sigma_i \rangle}{\partial T}\right)^{-1}_{T=T_0} = e^{-\frac{2J_\mathrm{eff}}{k_\mathrm{B}T_0}}\frac{k_\mathrm{B}T_0^2}{-\cosh(\beta h_\mathrm{eff}\mu_a)}\left[h_\mathrm{eff}\mu_a+\beta \mu_a\left(\frac{\partial h_\mathrm{eff}}{\partial \beta}\right)\right]^{-1}_{\beta=\frac{1}{k_\mathrm{B}T_0}}. \label{dT_eff}
\end{equation}
Eq.~(\ref{OP_eff}) and Eq.~(\ref{dT_eff}) are the same as Eq.~(\ref{OP_general}) and Eq.~(\ref{dT_general}), respectively. These two different representations can be used to verify the results obtained from using the other method.

For the Ising diamond chain model defined in Eq.~(\ref{diamond}) and illustrated in Fig.~1\boldfont{e}, 
\begin{eqnarray}
{\boxedpph}&=&2\cosh(4\beta J_1 + 2\beta h \mu_b)e^{\beta J_2}+2e^{-\beta J_2},\nonumber \\
{\boxedmmh}&=&2\cosh(4\beta J_1 - 2\beta h \mu_b)e^{\beta J_2}+2e^{-\beta J_2}, \\
{\boxedpmh}&=&{\boxedmph}=2\cosh(2\beta h \mu_b)e^{\beta J_2}+2e^{-\beta J_2}.\nonumber
\end{eqnarray} 
For $J_2=0$, the Ising diamond chain does not have geometric frustration, as shown in Fig.~1\boldfont{f}, 
\begin{eqnarray}
{\boxedpph}&=&[2\cosh(2\beta J_1 + \beta h\mu_b)]^2,\nonumber \\
{\boxedmmh}&=&[2\cosh(2\beta J_1 - \beta h\mu_b)]^2, \\
{\boxedpmh}&=&{\boxedmph}=[2\cosh(\beta h\mu_b)]^2.\nonumber
\end{eqnarray} 
\begin{equation}
\begin{split}
    \delta T = e^{-\frac{2J}{k_\mathrm{B}T_0}}\frac{2k_\mathrm{B}T_0^2 \boxedpmh}{-\Upsilon_+}\biggl[h\mu_a &+ (2 J_1+ h \mu_b)\tanh(2\beta J_1+\beta h \mu_b) \\
    &- (2 J_1- h \mu_b)\tanh(2\beta J_1-\beta h \mu_b)\biggr]^{-1}_{\beta=\frac{1}{k_\mathrm{B}T_0}}. 
\end{split}
    \label{dT_2}
\end{equation}

To compare with the previously reported results~\cite{016_Strecka_book_chapter}, the following transformation is needed:
\begin{equation}
    J \to J S^2, \;\; J_1 \to J_1 S^2, \;\; J_2 \to J_2 S^2, \;\; \mu_a=\mu_b=S,
\end{equation}
where $S=1/2$, as done for all the results presented in this manuscript.

\section{The Magnetization}
\red{It is noteworthy that the magnetization $m=\mu_a \langle \sigma_i \rangle + \mu_b \sum_k\langle s_{i,k} \rangle$ offers less clear evidence of the CPT at $T_0$. It was shown that for $J_2/J_1=1.95$ and $J=0$, no traces of the pseudo-transition could be found in the density plot of $m$ [see Fig.~\ref{Fig:T-h-m}(a)], while the pseudo-transition could be clearly seen in the respective density plots of the entropy, susceptibility, and specific heat~\cite{016_Strecka_book_chapter}. This insensitivity of $m$ can be explained by the fact that $J_2/J_1=1.95$ is near the FRI-FRU phase boundary where $m$ is almost the same on both sides with exactly two spins up and one down per unit cell in the ground state as shown in Fig.~\ref{Fig:zeroT}. This feature remains unchanged for $J>0$ [see Fig.~\ref{Fig:T-h-m}(b)]. When moving away from the FRI-FRU phase boundary, $m$ becomes a better indicator of the CPT; for example, for $J_2=0$, the hidden frustration involves the SPP phase and the CPT can be revealed in the density plot of $m$ except for the weak $h$ region [see Fig.~\ref{Fig:T-h-m}(d)]. The comparison between $m$ and $\langle \sigma_i \rangle$ the OP will be addressed in more details in subsequent publications~\cite{Yin_Ising_III}.}

\newpage
\begin{figure}[t]
    \begin{center}
\includegraphics[width=0.6\columnwidth,clip=true,angle=0]{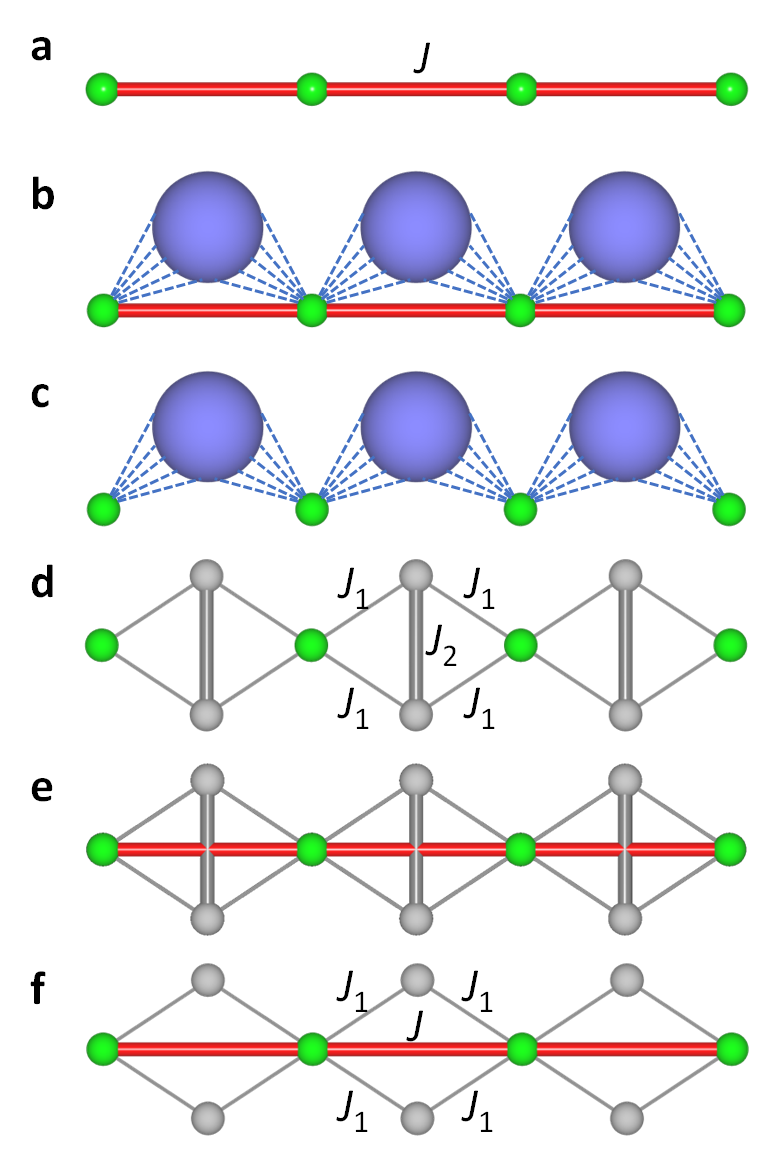}
    \end{center}
\caption{\figtextbf{Schematic diagrams of decorated Ising chains.} (a) The original ordinary Ising chain~\cite{Ising1925}, which consists of the
Ising spins (green balls) with the ferromagnetic interaction $J>0$ (red bonds). (b) The decoration
consists of arbitrary finite-size sublattices (big blue balls) which are coupled to the ordinary chain by the Ising-type interactions (dotted lines). (c) Decorated Ising chain with the $J$ bonds neglected~\cite{005_Galisova_PRE_15_double-tetrahedral-chain,
007_Torrico_PRA_16_Ising-XYZ-diamond-chain,
009_review_Souza_SSC_18_Ising-XYZ-diamond-chain_double-tetrahedral-chain-spin-electron,
010_Carvalho_JMMM_18_Ising-XYZ-diamond-chain_quantum-entanglement,
011_Rojas_BJP_20_Ising-Heisenberg-tetrahedral_diamond,
013_Rojas_PRE_19_previous_4_models,
014_Rojas_JPC_20_Ising–Heisenberg_spin-1-double-tetrahedral-chain,
015_Strecka_APPA_20_Ising-diamond-chain,
015-7_Canova_CzechoslovakJP_04_Ising–Heisenberg_diamond_chain,
015-8_Canova_JPC_06_Ising–Heisenberg-spin-S-diamond-chain,
017_Krokhmalskii_PA_21_3-previous-chains_effective_model,
016_Strecka_book_chapter}. (d) The Ising diamond chain for $J=0$~\cite{016_Strecka_book_chapter,015_Strecka_APPA_20_Ising-diamond-chain}, in which the decorated Ising spins (grey balls) utilize the antiferromagnetic interactions $J_1<0$ and $J_2<0$ (grey bonds) to form triangles, yielding  geometric frustration. (e) The Ising diamond chain with ferromagnetic $J>0$ studied here. (f) The Ising diamond chain without geometric frustration for $J_2=0$.}
\label{Fig:structure}
\end{figure}

\newpage
\begin{figure}[t]
    \begin{center}
        \subfigure[][]{
\includegraphics[width=0.6\columnwidth,clip=true,angle=0]{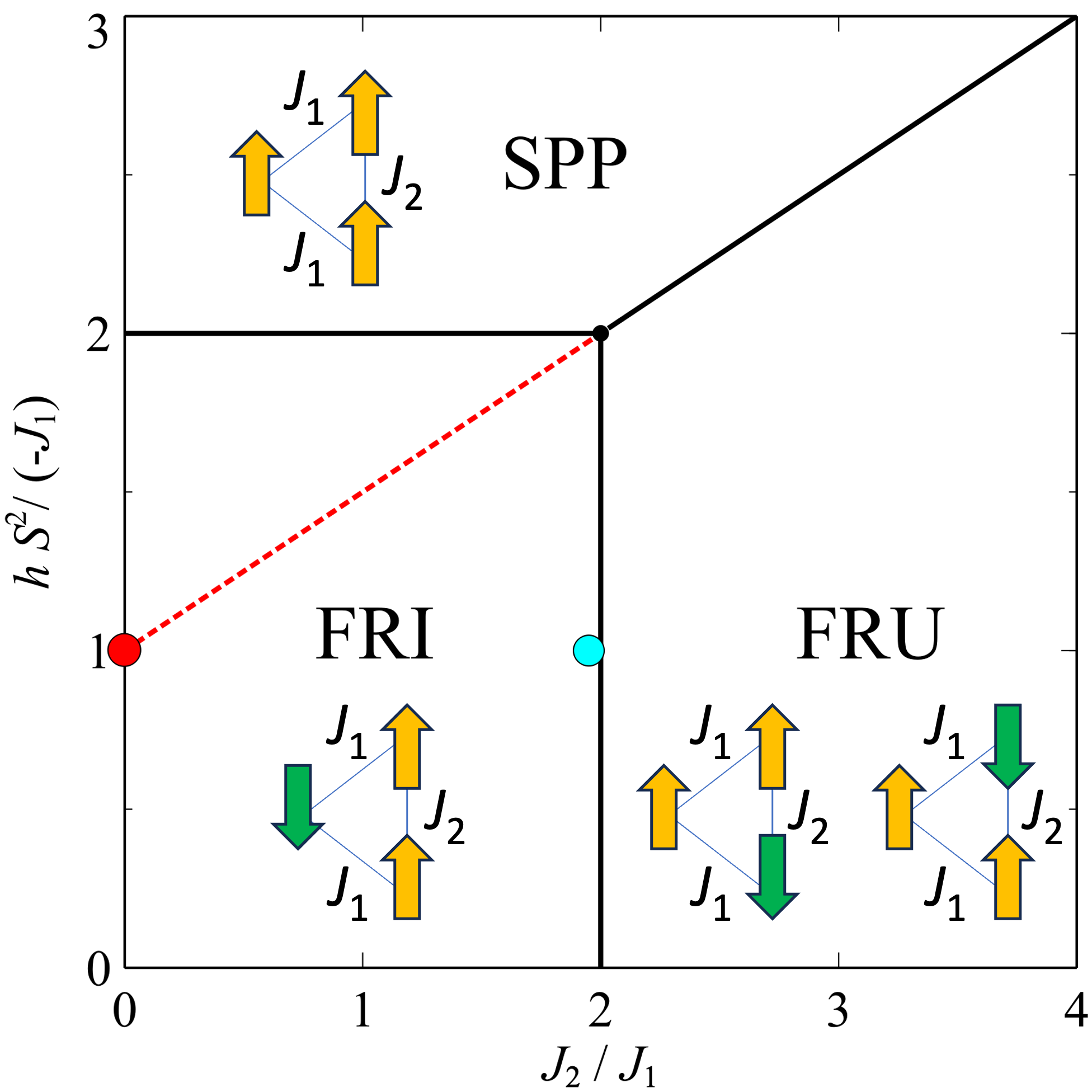}
        }
    \end{center}
\caption{\figtextbf{The ground-state phase diagram} of the Ising diamond chain in the $J_2/J_1-h/J_1$ plane. Following Ref.~\cite{016_Strecka_book_chapter}, we use the same notation for individual ground states to facilitate the comparison: FRI - the ferrimagnetic phase, FRU - the frustrated phase, SPP - the saturated paramagnetic phase. The focus of Ref.~\cite{016_Strecka_book_chapter} was the case with strong geometric frustration (the cyan circle for $J_2/J_1=1.95$ near the FRI $-$ FRU boundary) as the FRU phase has the degeneracy of two per unit cell. Here we also study the case without geometric frustration (the red circle for $J_2=0$), which is located deep in the FRI phase, far away from the other two phases, but right on the extended line of the FRU $-$ SPP boundary (red dashed line) implying a \emph{hidden}  excited state with remarkably higher degeneracy than the FRI ground state. $J_1=-S^2$ and $\mu_a=\mu_b=S$ where $S=1/2$.}
\label{Fig:zeroT}
\end{figure}

\newpage
\begin{figure}[t]
    \begin{center}
        \subfigure[][]{
\includegraphics[width=0.48\columnwidth,clip=true,angle=0]{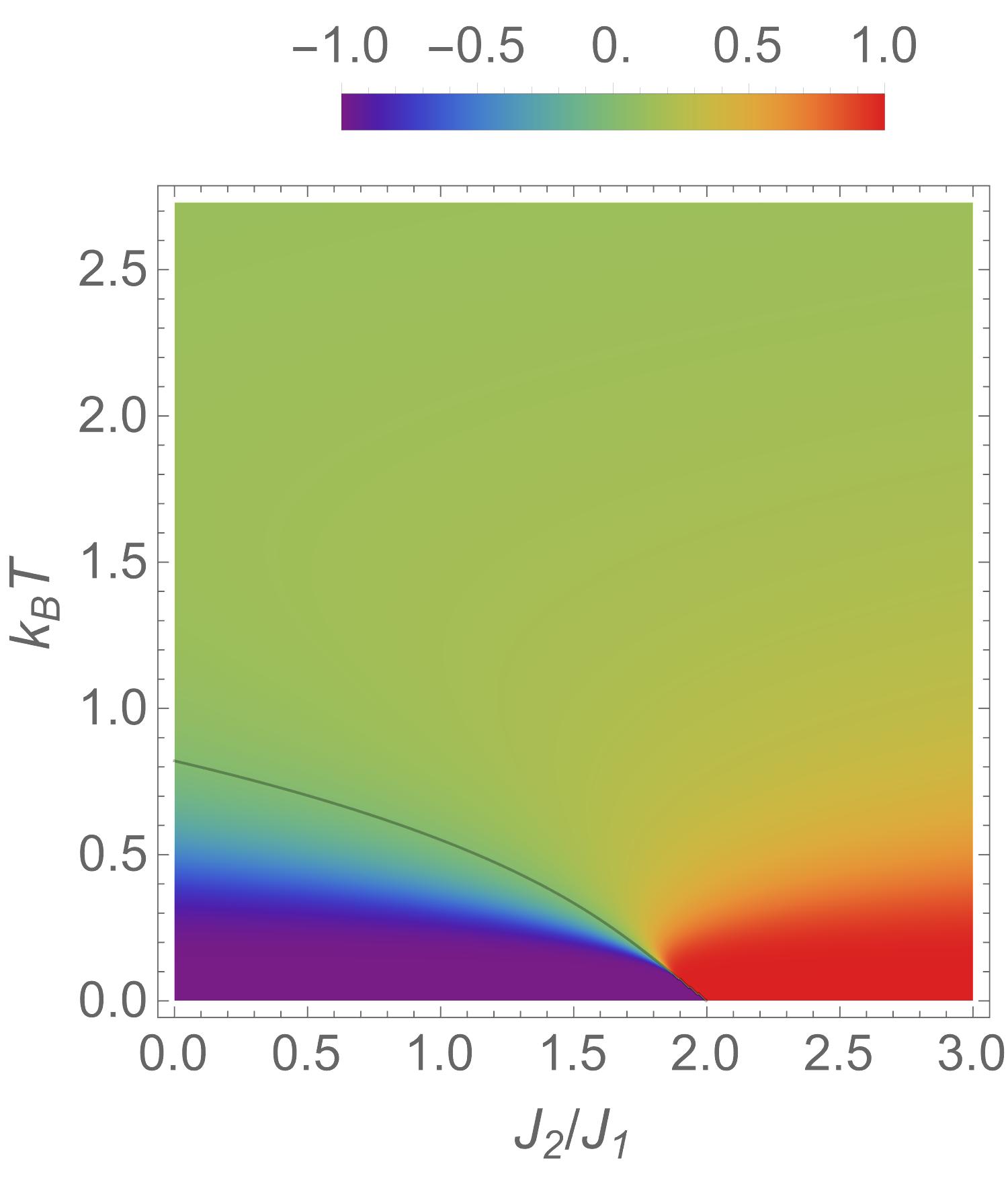}
        }
        \subfigure[][]{
\includegraphics[width=0.48\columnwidth,clip=true,angle=0]{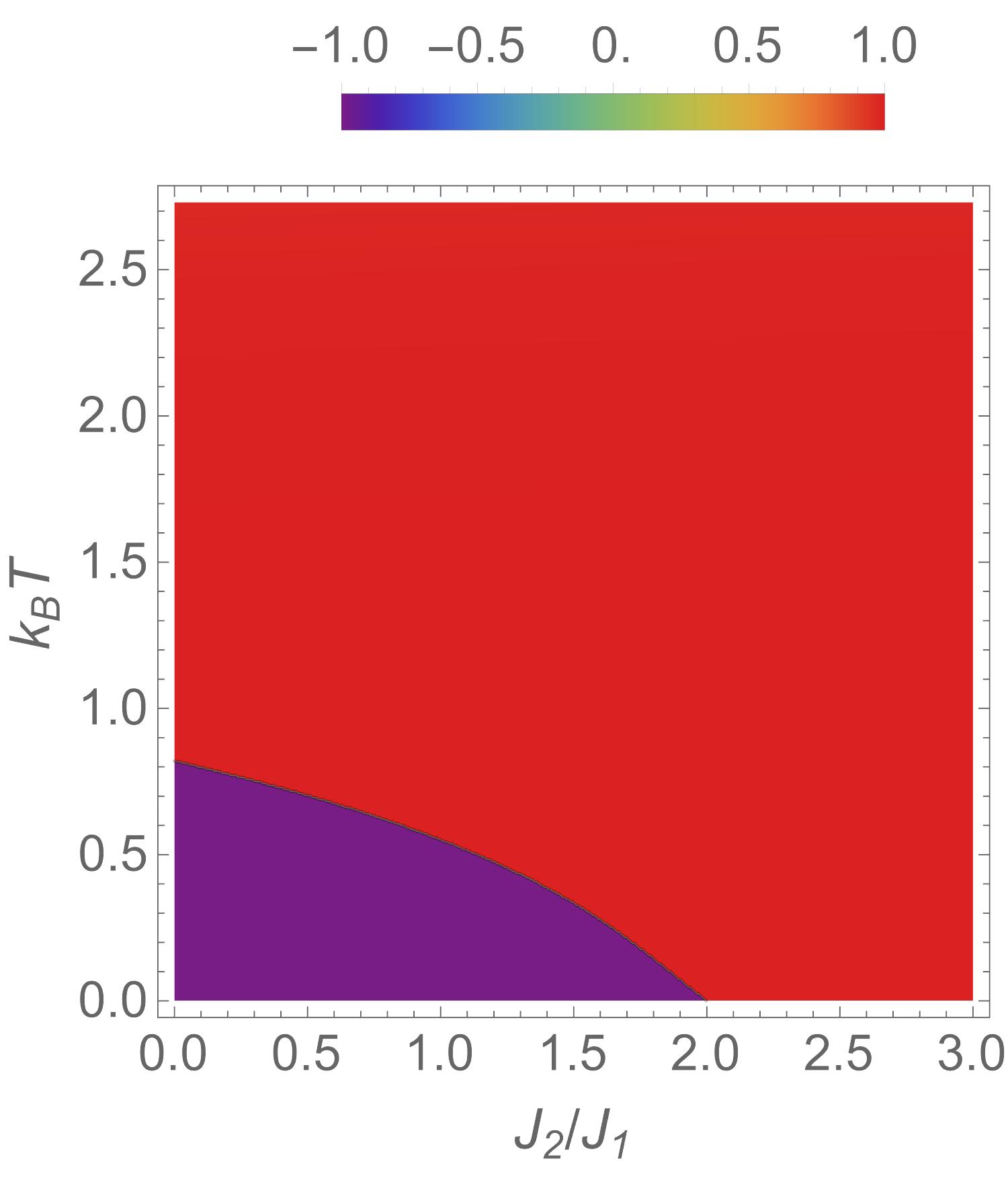}
        }
    \end{center}
\caption{\figtextbf{The $T - J_2$ phase diagrams from the density plot of the order parameter $\langle\sigma_i\rangle$} for (a) $J=0$ and (b) $J=20S^2$. The black line is the phase boundary where $\langle\sigma_i\rangle=0$. $J_1=-S^2$, $h=1$, and $\mu_a=\mu_b=S$ where $S=1/2$.}
\label{Fig:T-J2}
\end{figure}

\newpage
\begin{figure}[t]
    \begin{center}
        \subfigure[][]{
\includegraphics[width=0.48\columnwidth,clip=true,angle=0]{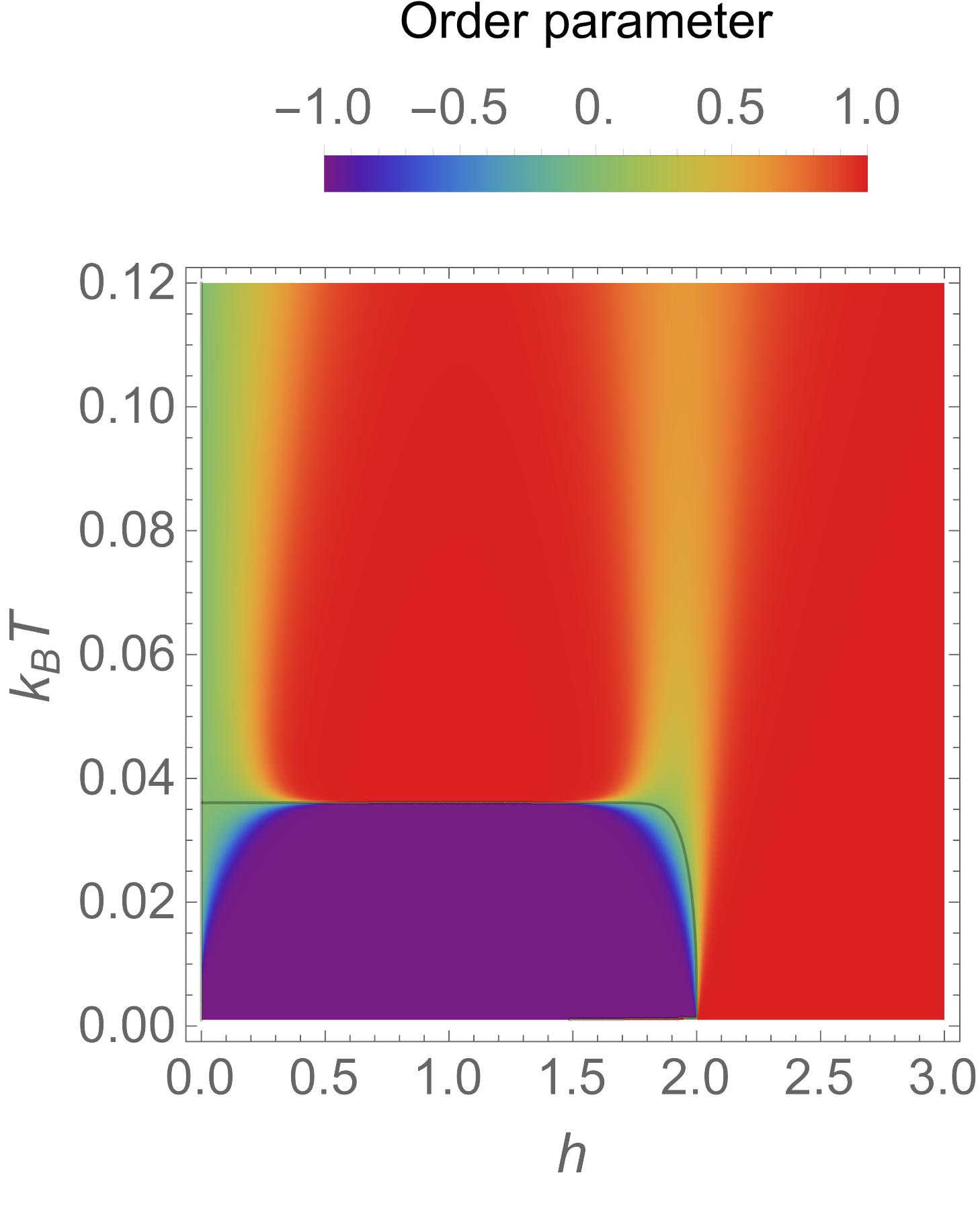}
        }
        \subfigure[][]{
\includegraphics[width=0.48\columnwidth,clip=true,angle=0]{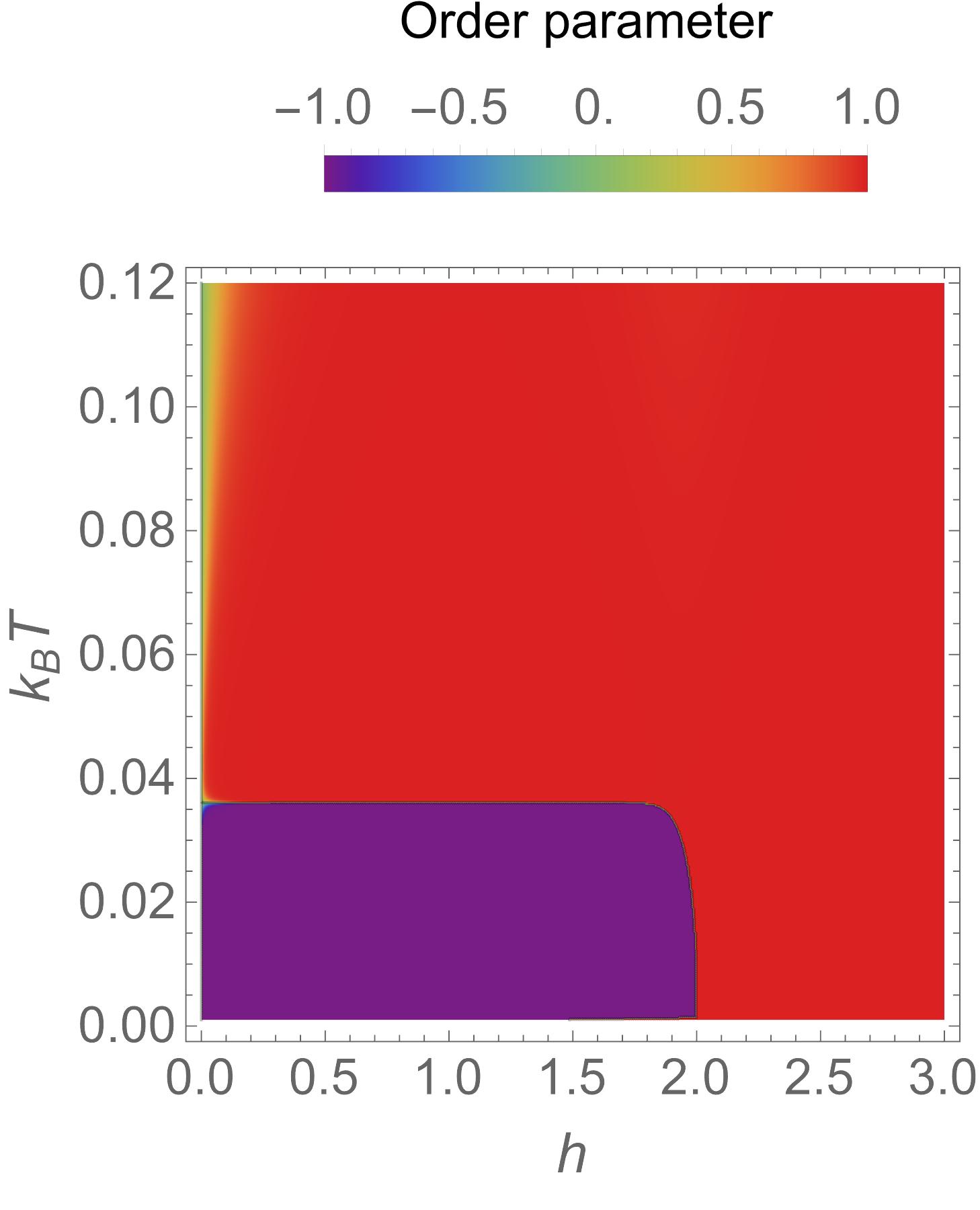}
        }
        \subfigure[][]{
\includegraphics[width=0.48\columnwidth,clip=true,angle=0]{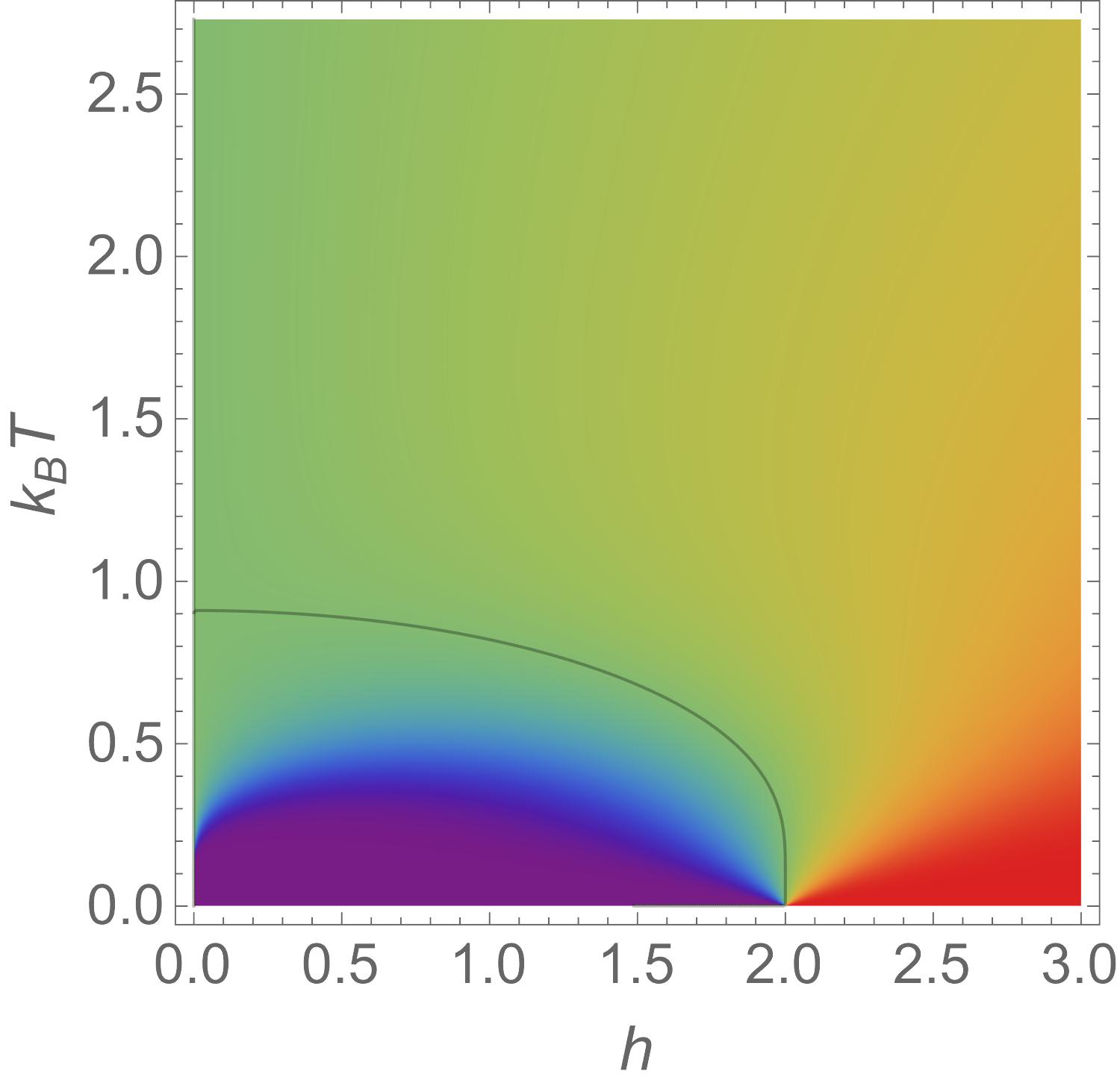}
        }
        \subfigure[][]{
\includegraphics[width=0.48\columnwidth,clip=true,angle=0]{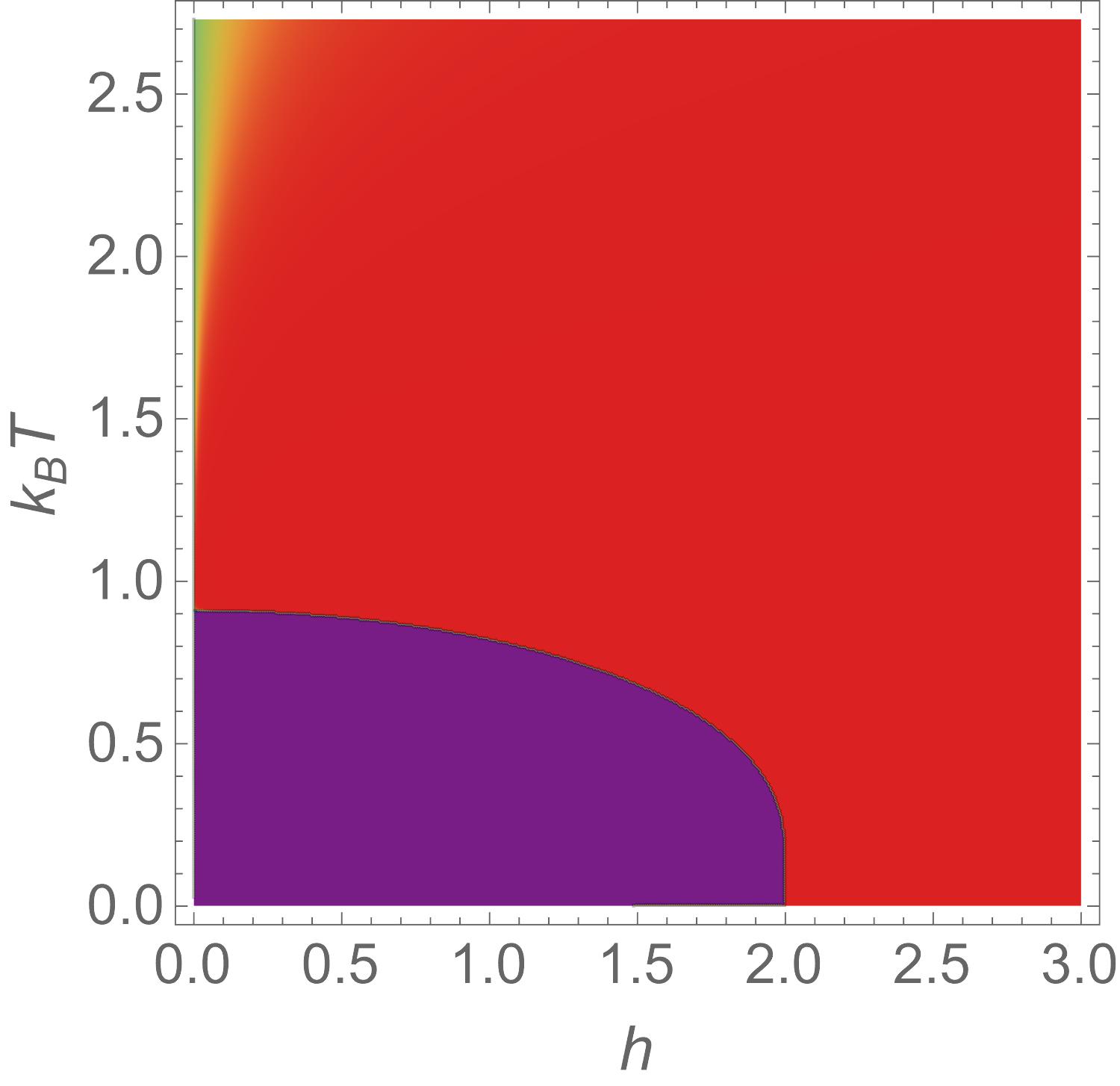}
        }
    \end{center}
\caption{\figtextbf{The $T - h$ phase diagrams from the density plot of the order parameter $\langle\sigma_i\rangle$} for (a) $J_2/J_1=1.95$ and $J=0$, (b) $J_2/J_1=1.95$ and $J=0.4S^2$, (c) $J_2/J_1=0$ and $J=0$, (d) $J_2/J_1=0$ and $J=20S^2$. The black line is the phase boundary where $\langle\sigma_i\rangle=0$. $J_1=-S^2$ and $\mu_a=\mu_b=S$ where $S=1/2$.}
\label{Fig:T-h}
\end{figure}

\newpage
\begin{figure}[t]
    \begin{center}
        \subfigure[][]{
\includegraphics[width=0.48\columnwidth,clip=true,angle=0]{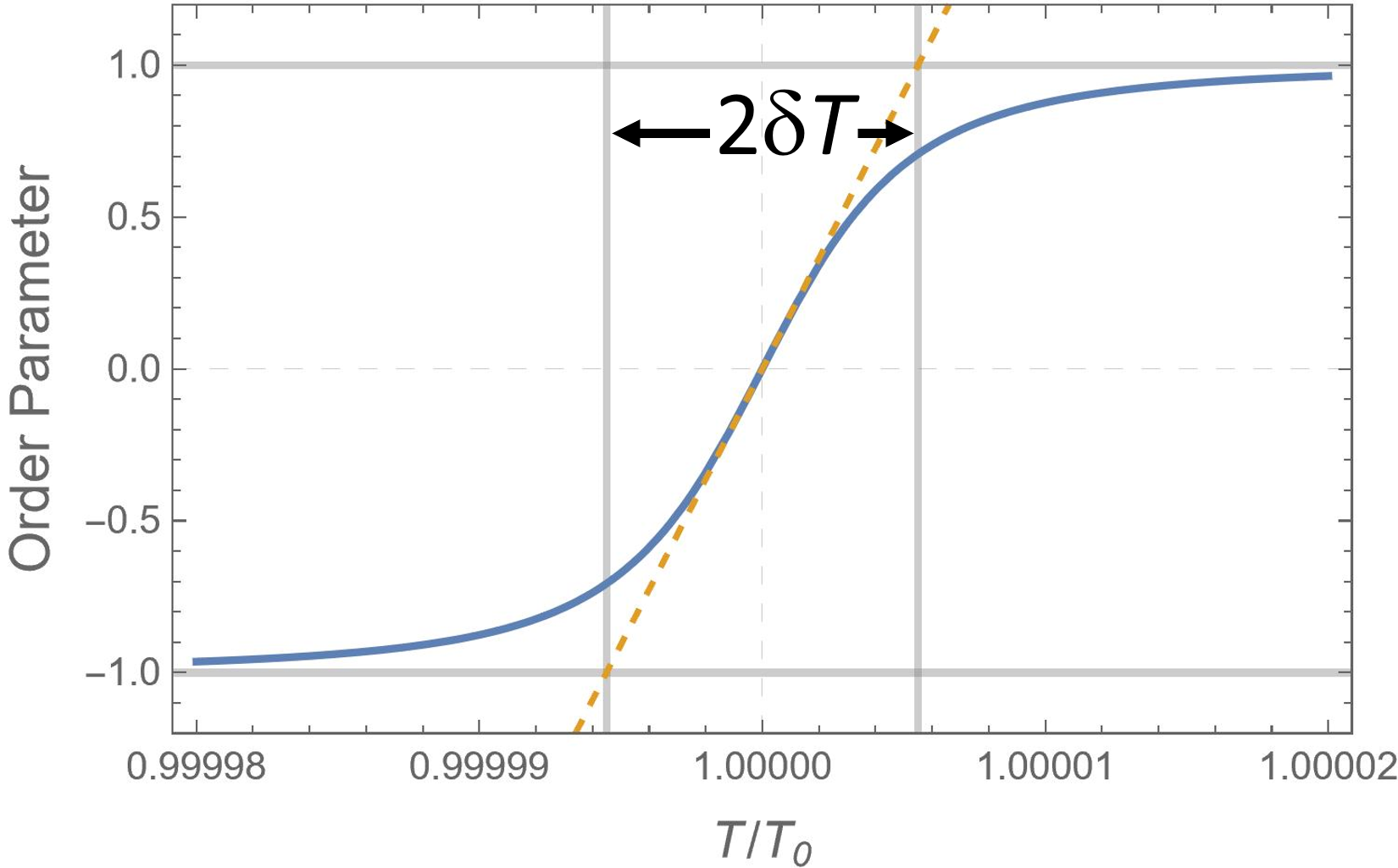}
        }
        \subfigure[][]{
\includegraphics[width=0.48\columnwidth,clip=true,angle=0]{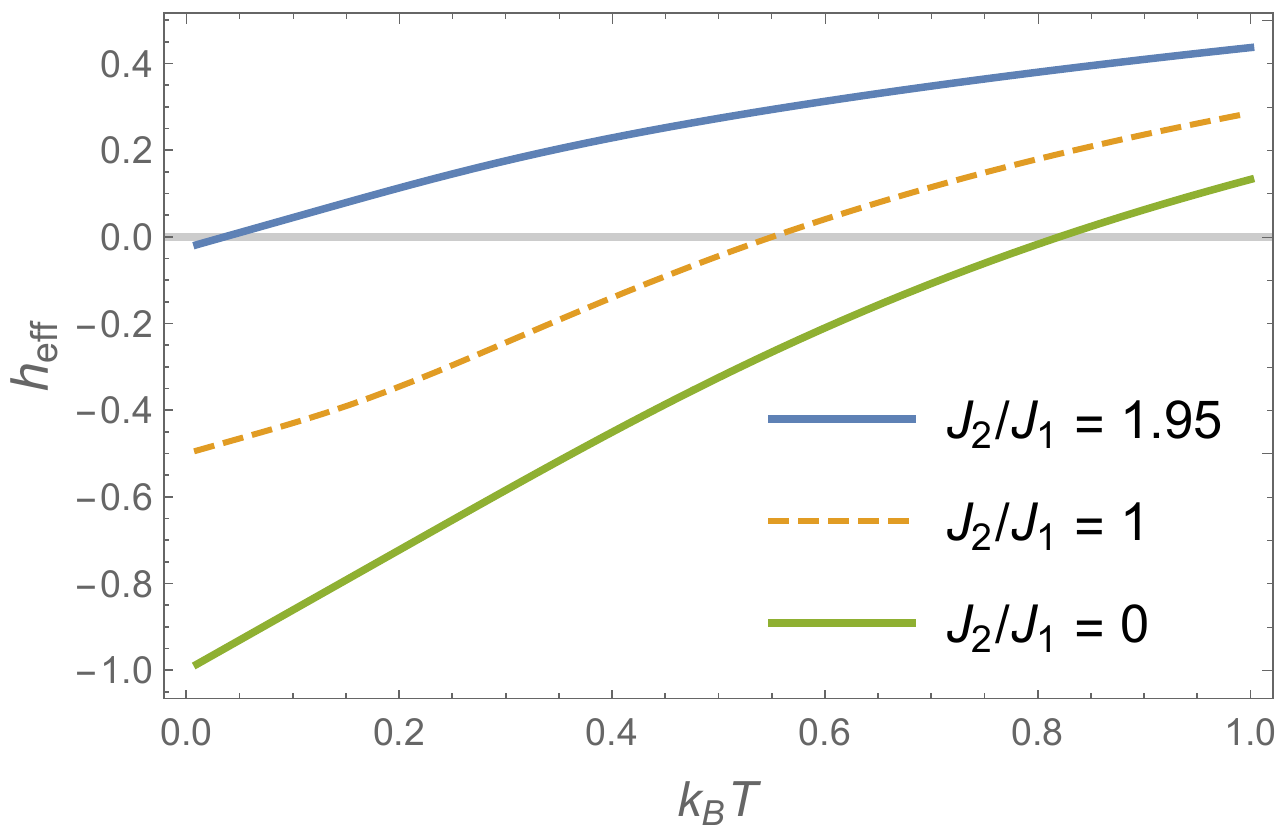}
        }
        \subfigure[][]{
\includegraphics[width=0.48\columnwidth,clip=true,angle=0]{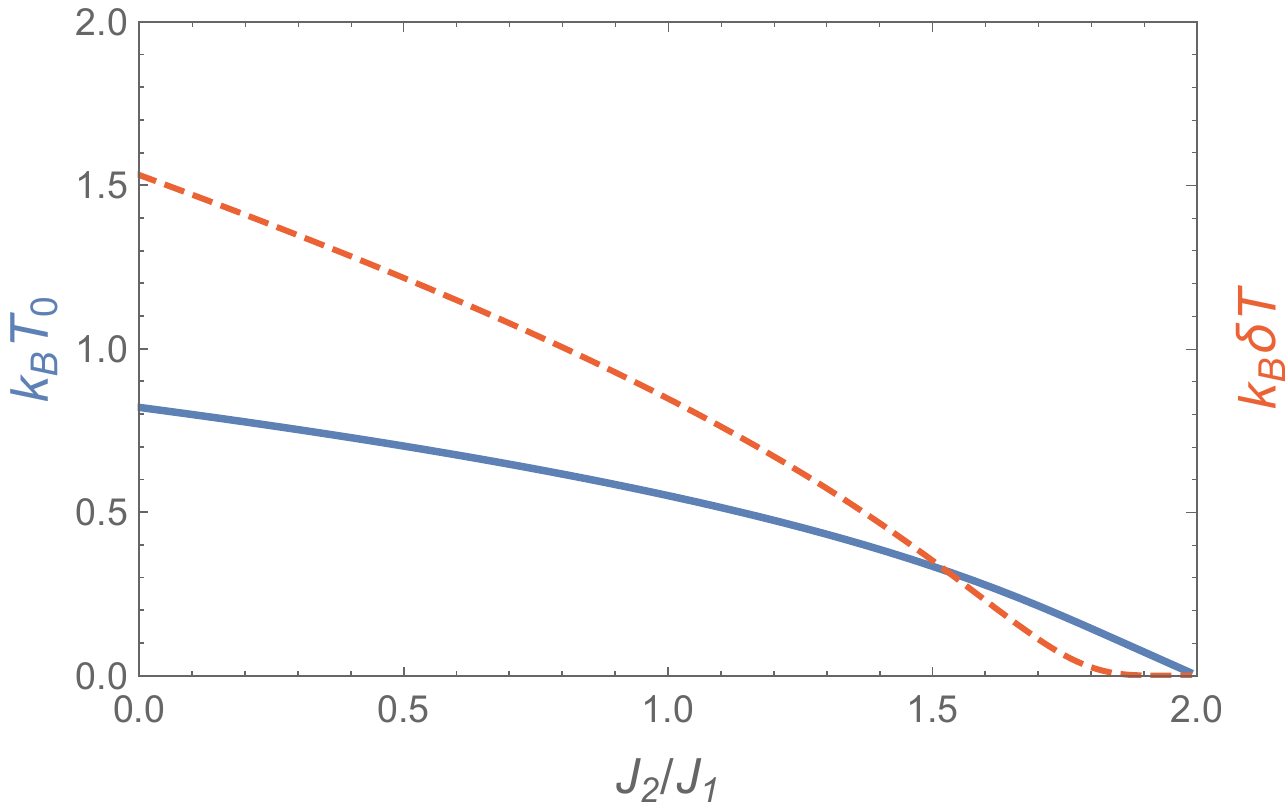}
        }
        \subfigure[][]{
\includegraphics[width=0.48\columnwidth,clip=true,angle=0]{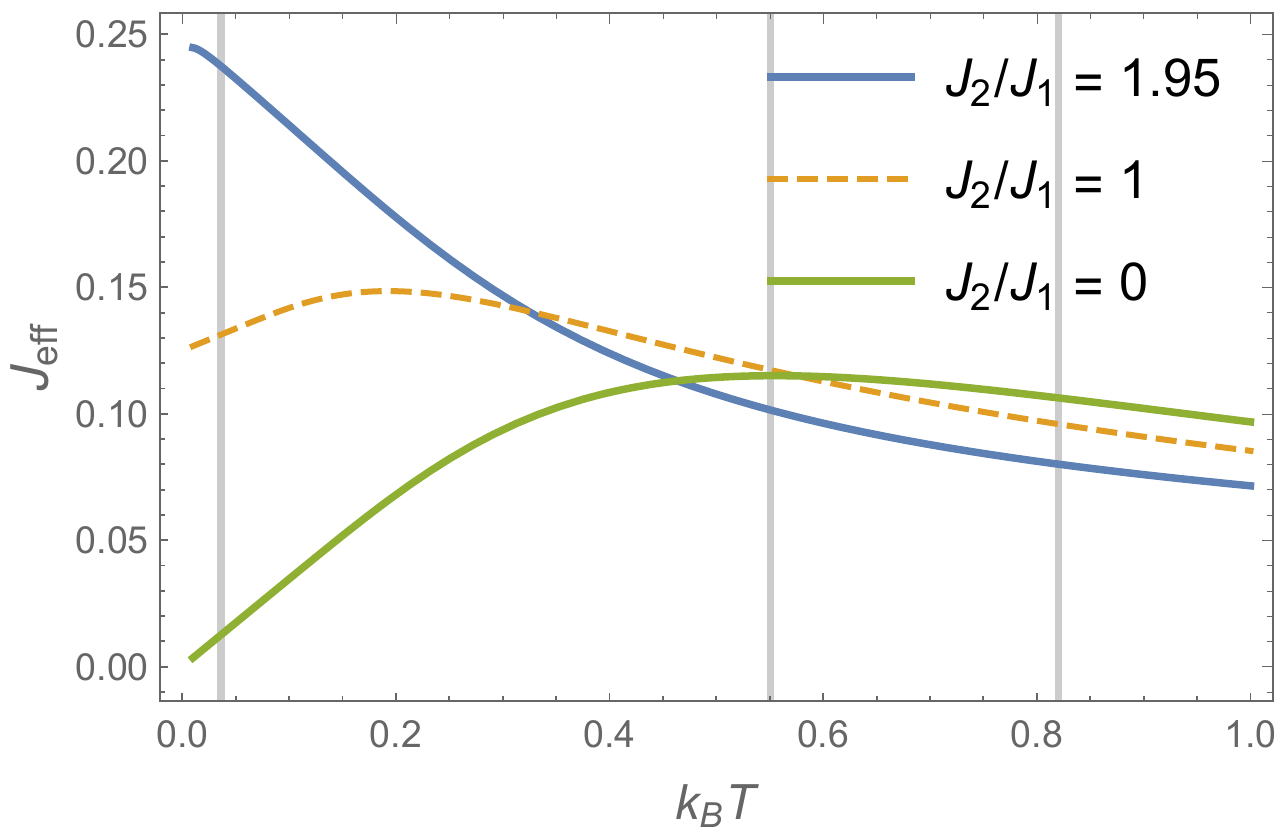}
        }
        \subfigure[][]{
\includegraphics[width=0.48\columnwidth,clip=true,angle=0]{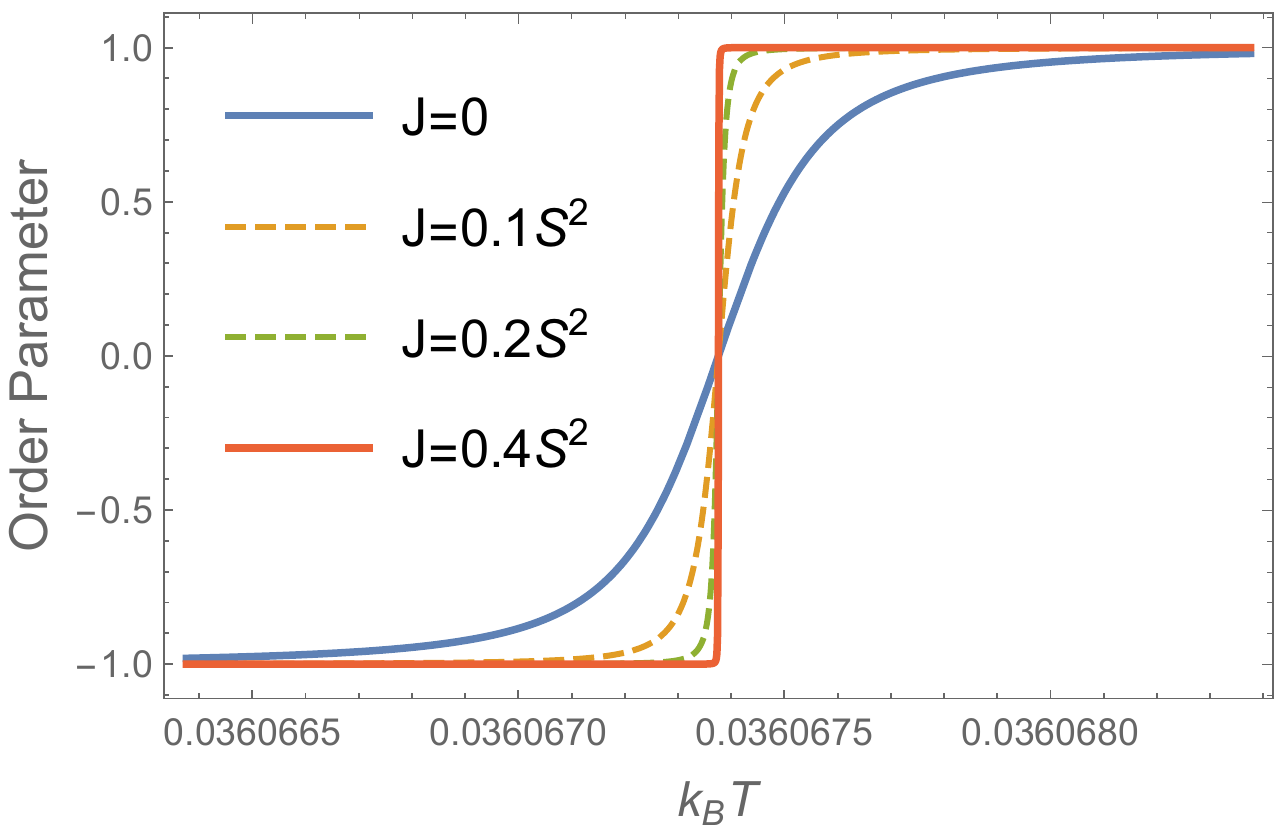}
        }
        \subfigure[][]{
\includegraphics[width=0.48\columnwidth,clip=true,angle=0]{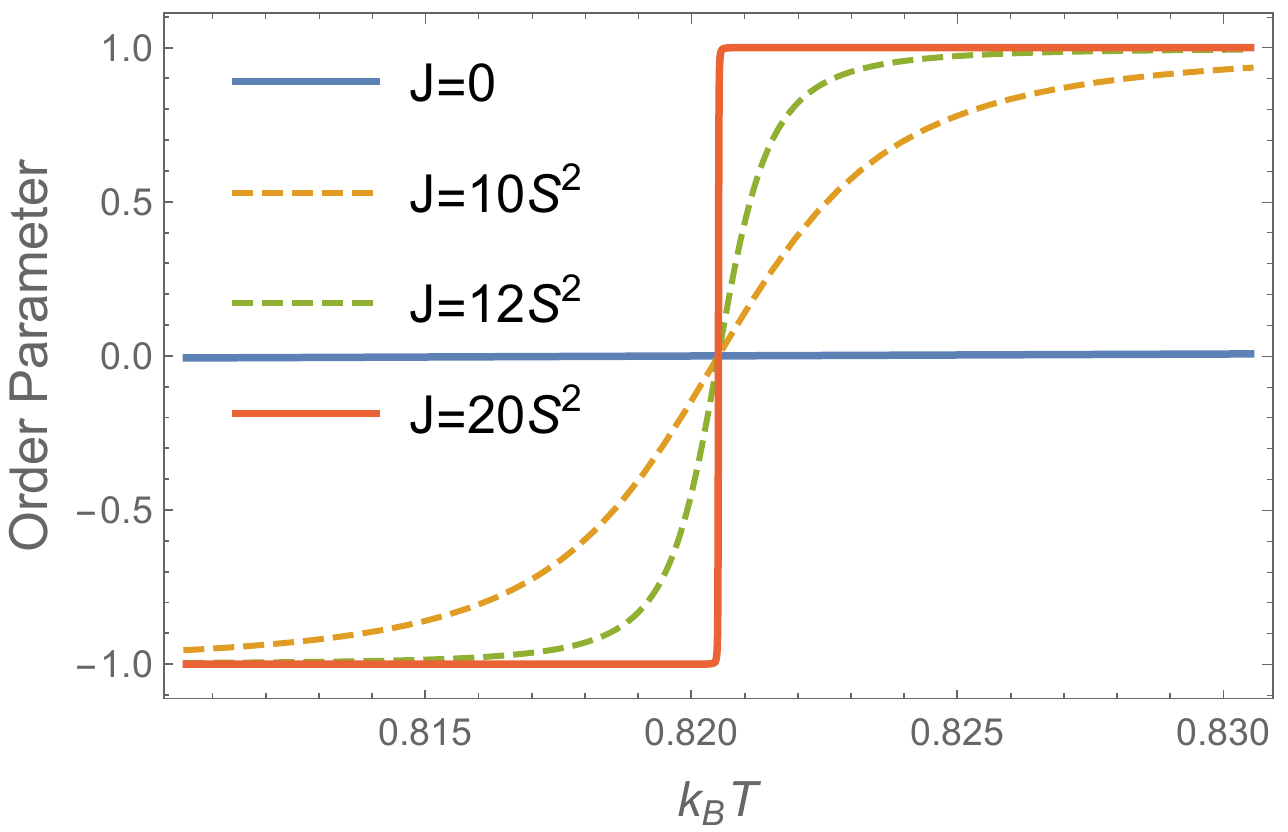}
        }
    \end{center}
\caption{\figtextbf{Order parameter $\langle \sigma_i\rangle$.} (a) The definition of the crossover width $2\delta T = 2\left(\partial \langle \sigma_i \rangle/\partial T\right)^{-1}_{T=T_0}$. The $T$ dependence of (b) $h_\mathrm{eff}$ and (d) $J_\mathrm{eff}$ for $J=0$ (light grey grid lines mark where $T_0$ is for $J_2/J_1=1.95, 1, 0$). (c) The $J_2/J_1$ dependence of $T_0$ (blue solid line; left axis) and $\delta T$ (red dashed line; right axis) for $J=0$. (e) The order parameter as a function of $T$ for $J_2/J_2=1.95$. (f) The order parameter as a function of $T$ for $J_2=0$. $J_1=-S^2$, $h=1$, and $\mu_a=\mu_b=S$ where $S=1/2$.}
\label{Fig:OP}
\end{figure}

\newpage
\begin{figure}[t]
    \begin{center}
        \subfigure[][]{
\includegraphics[width=0.48\columnwidth,clip=true,angle=0]{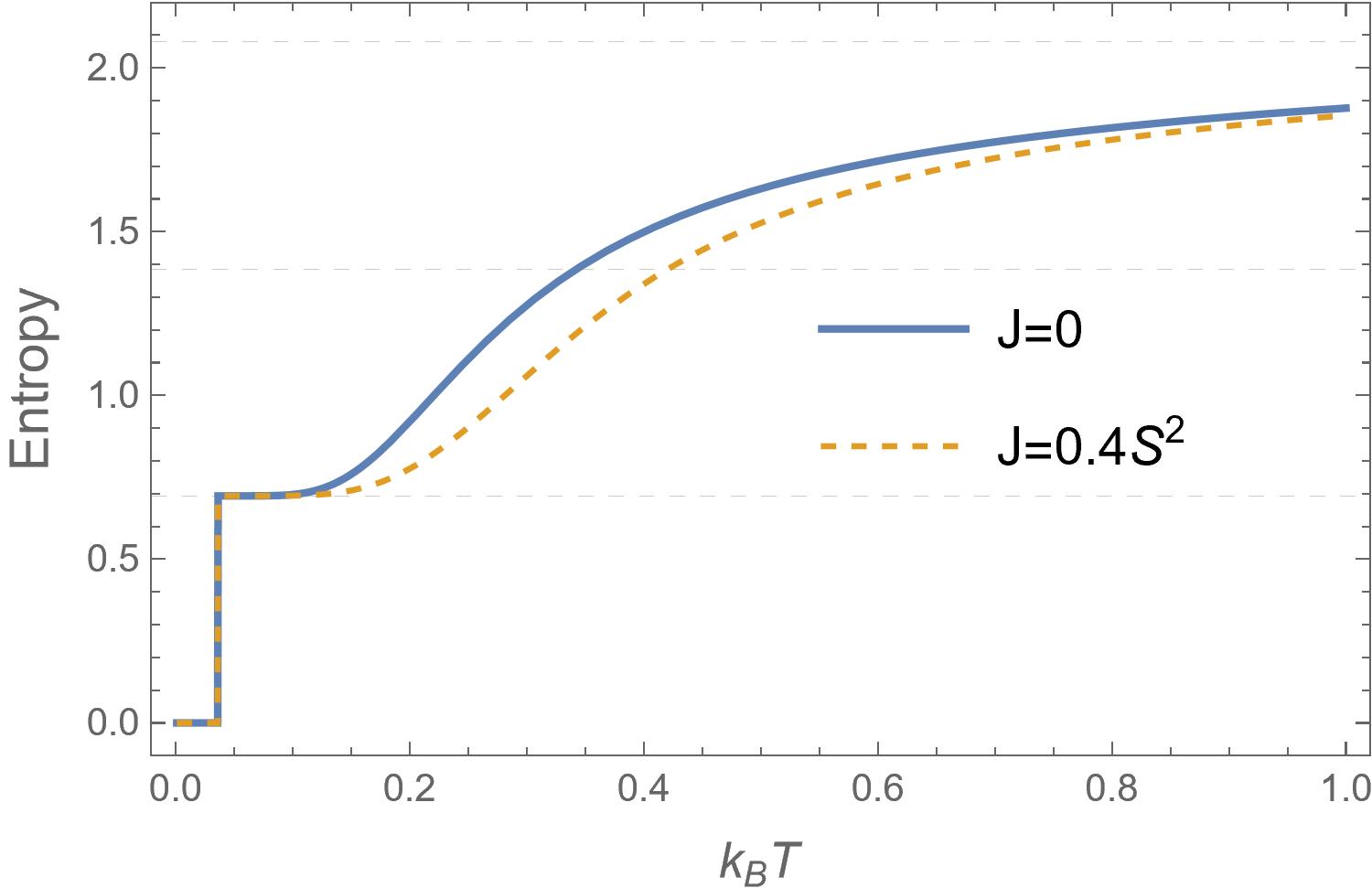}
        }
        \subfigure[][]{
\includegraphics[width=0.48\columnwidth,clip=true,angle=0]{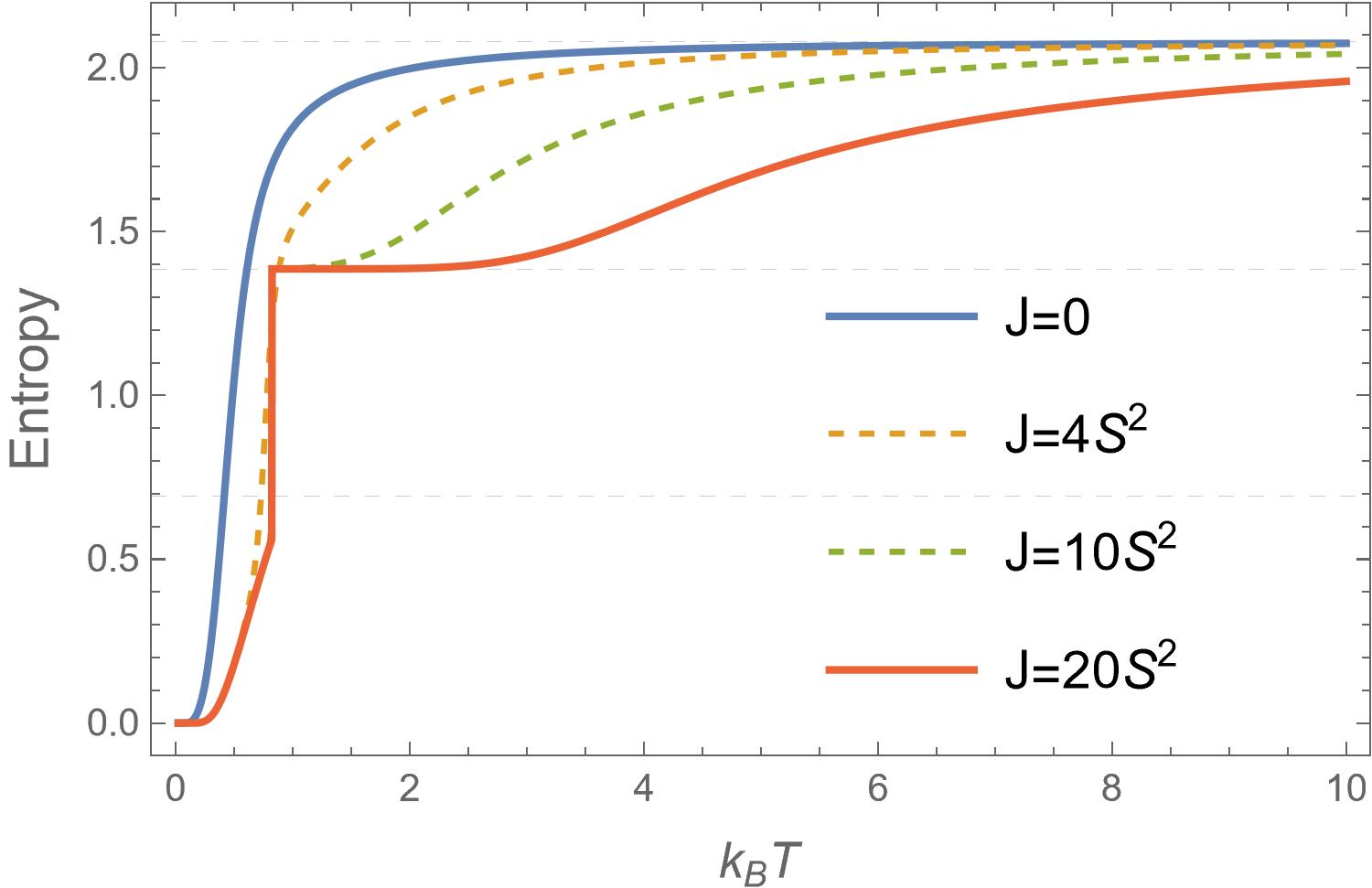}
        }
        \subfigure[][]{
\includegraphics[width=0.48\columnwidth,clip=true,angle=0]{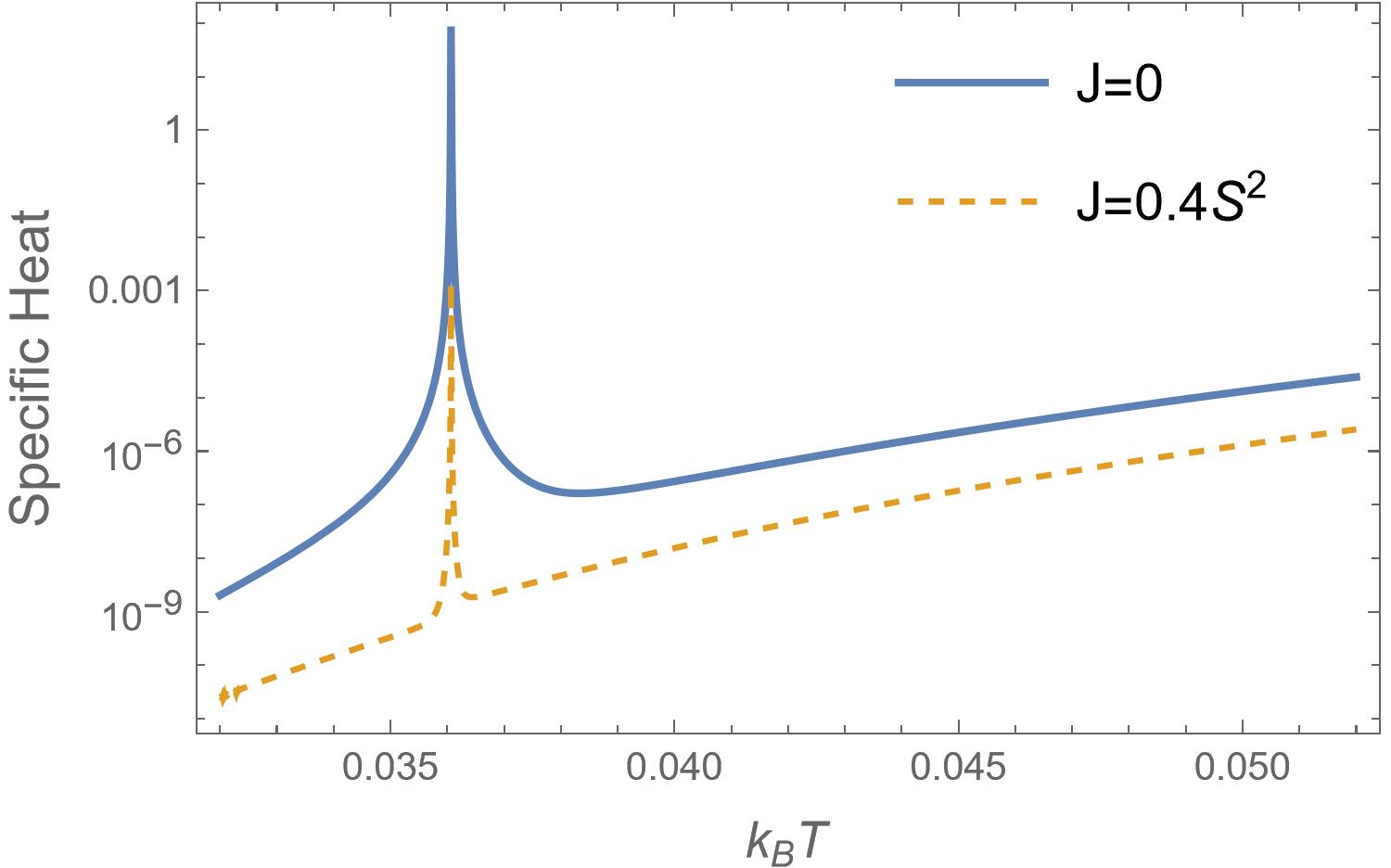}
        }
        \subfigure[][]{
\includegraphics[width=0.48\columnwidth,clip=true,angle=0]{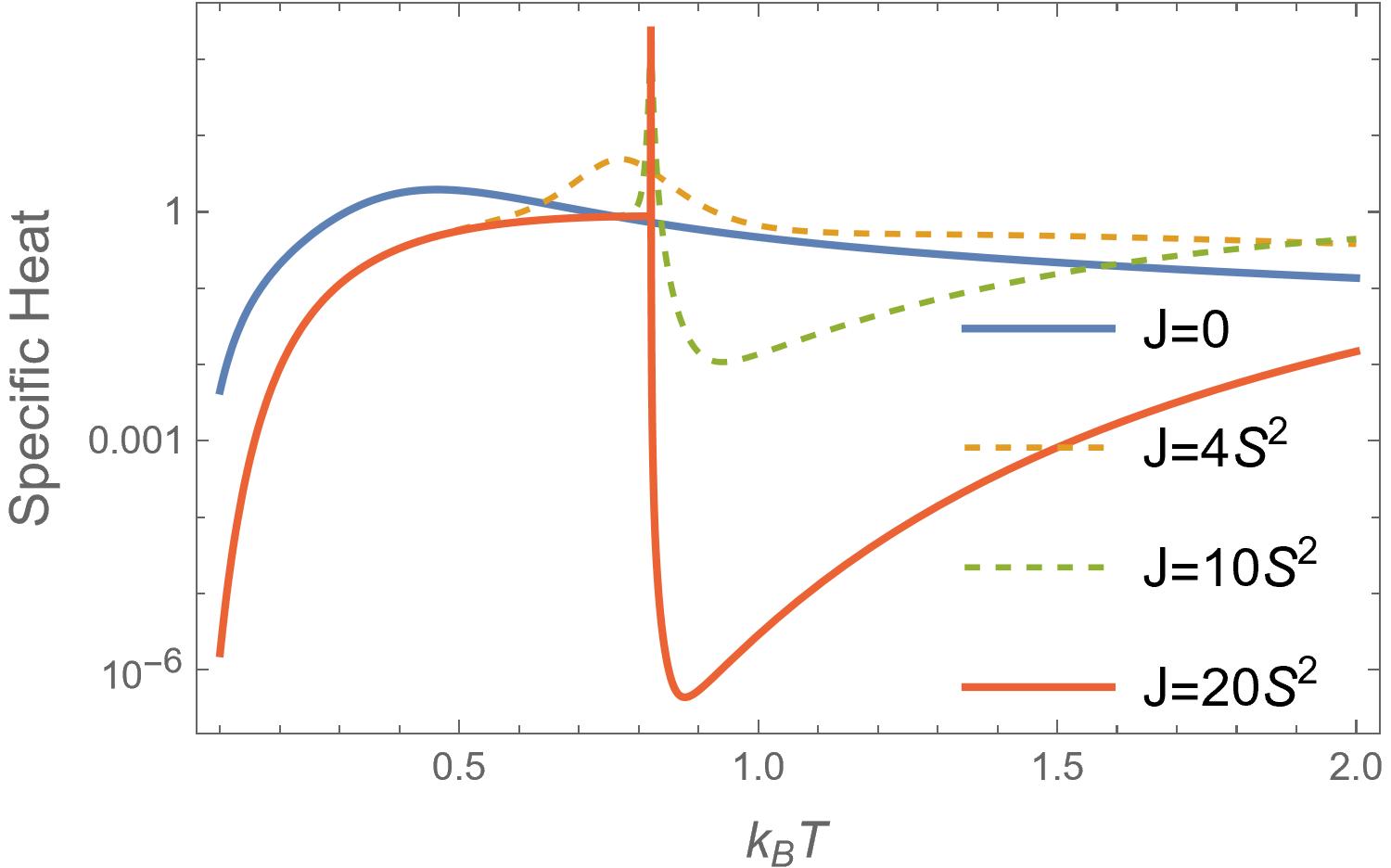}
        }
    \end{center}
\caption{\figtextbf{Thermodynamic properties.} Entropy for (a) $J_2/J_1=1.95$ and (b) $J_2/J_1=0$, where the three light grey dashed lines denote the values of $\ln2$, $2\ln2$, and $3\ln2$. Specific heat for (c) $J_2/J_1=1.95$ and (d) $J_2/J_1=0$. $J_1=-S^2$, $h=1$, and $\mu_a=\mu_b=S$ where $S=1/2$.}
\label{Fig:thermal}
\end{figure}

\newpage
\begin{figure}[t]
    \begin{center}
        \subfigure[][]{
\includegraphics[width=0.48\columnwidth,clip=true,angle=0]{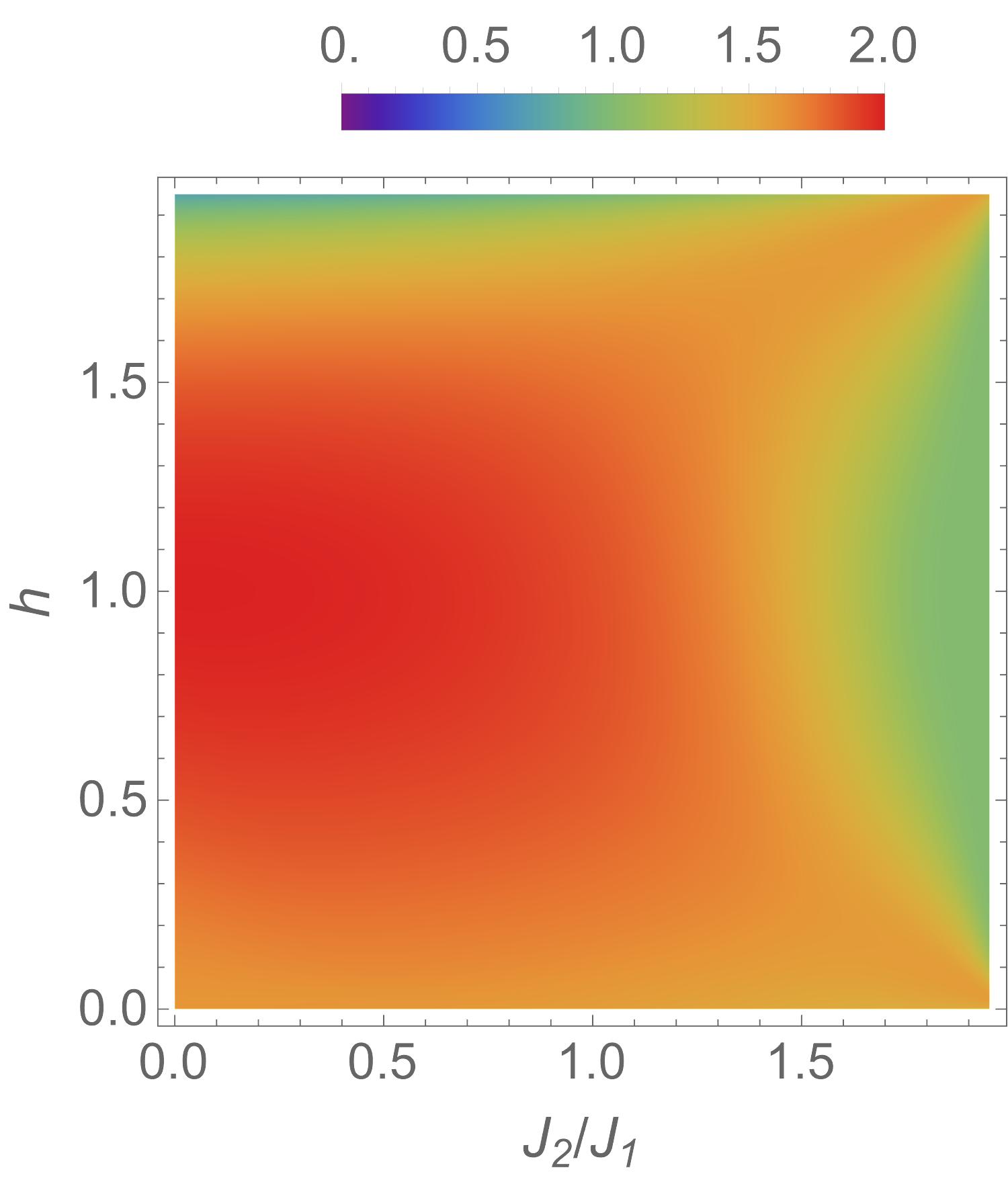}
        }
        \subfigure[][]{
\includegraphics[width=0.48\columnwidth,clip=true,angle=0]{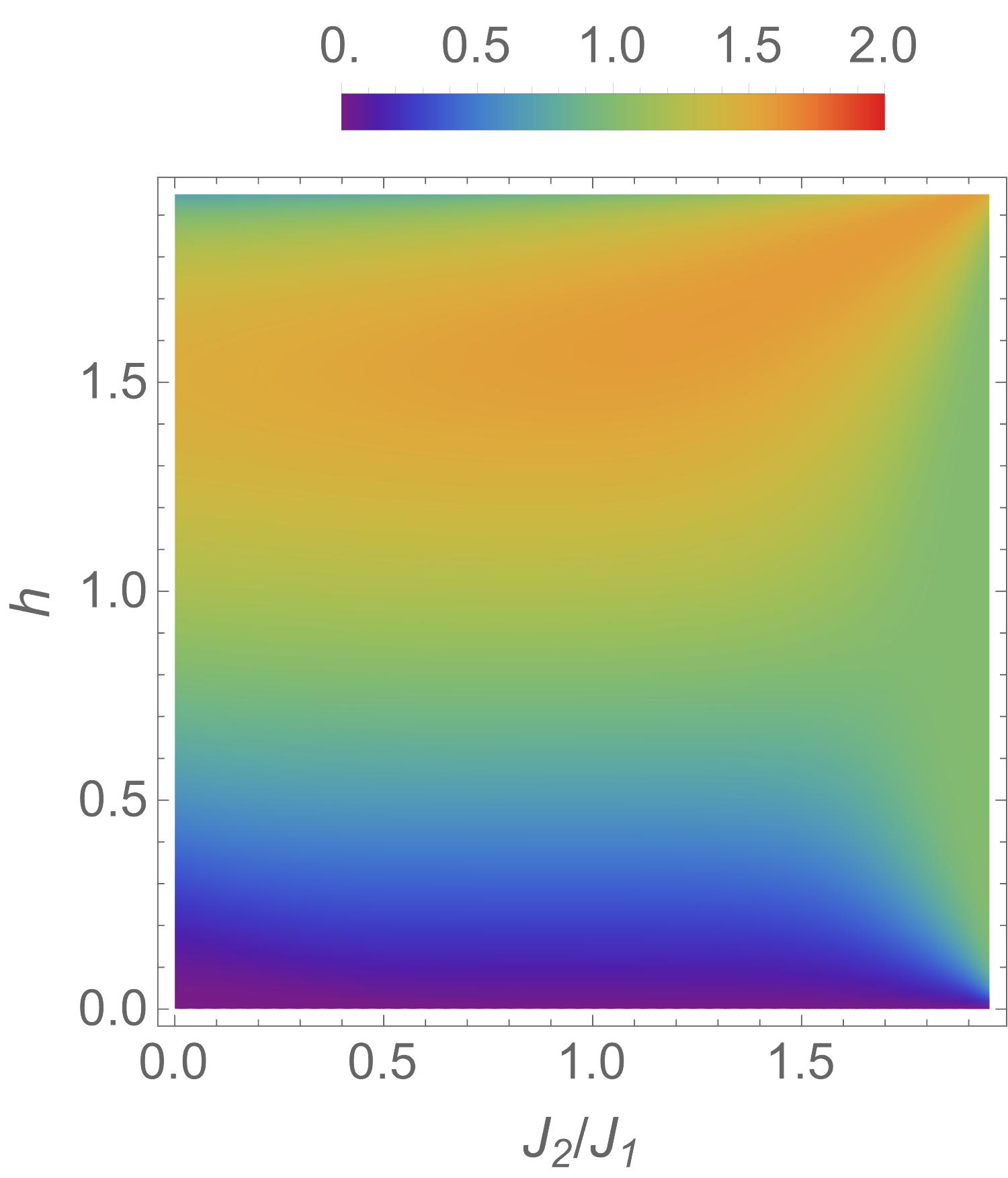}
        }
    \end{center}
\caption{\figtextbf{The $h-J_2$ density plots} of (a) the entropy right above $T_0$ and (b) the entropy jump at $T_0$ in the unit of $\ln 2$ per unit cell. $J=20S^2$, $J_1=-S^2$, and $\mu_a=\mu_b=S$ where $S=1/2$.}
\label{Fig:h-J2}
\end{figure}

\newpage
\begin{figure}[t]
    \begin{center}
        \subfigure[][]{
\includegraphics[width=0.48\columnwidth,clip=true,angle=0]{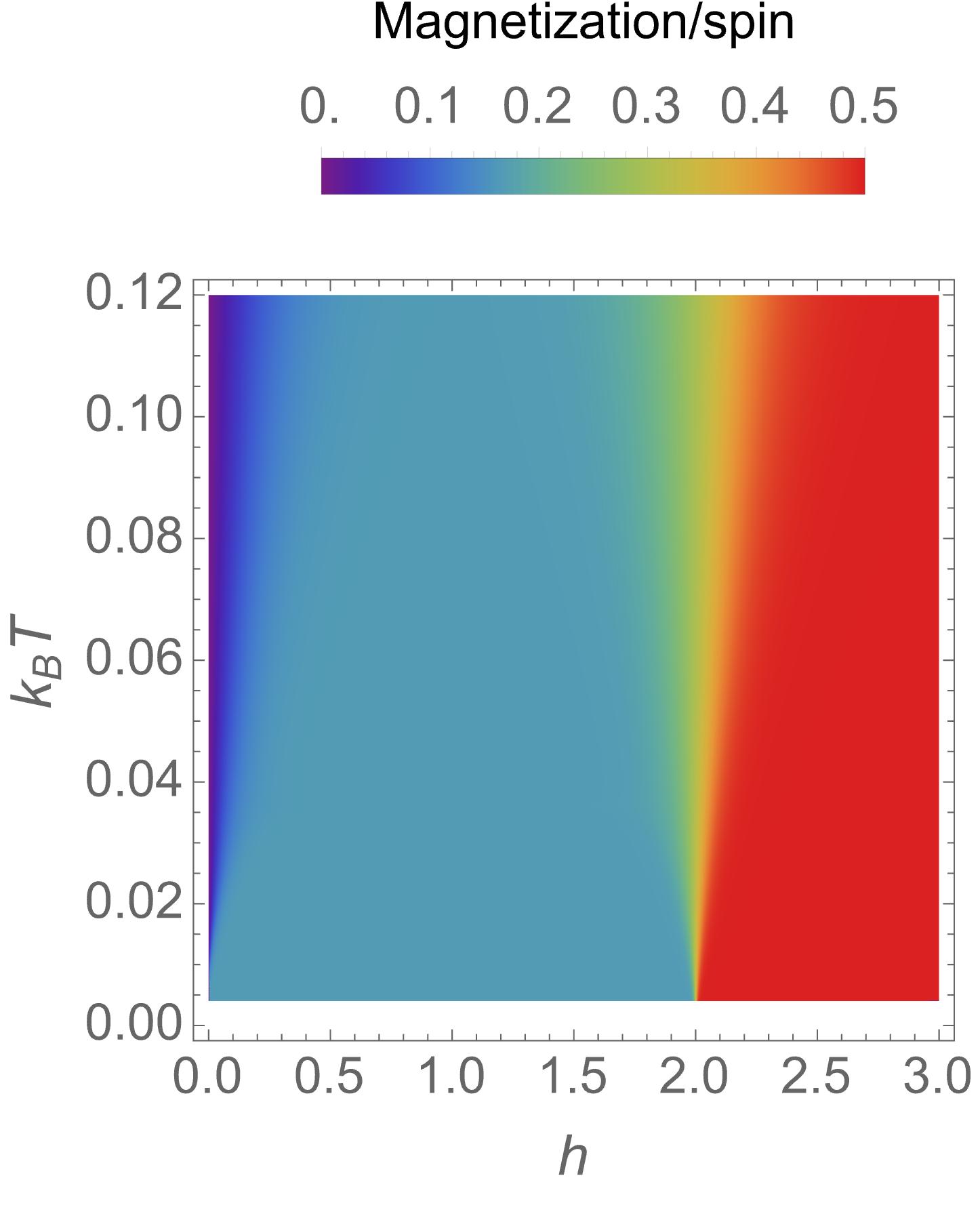}
        }
        \subfigure[][]{
\includegraphics[width=0.48\columnwidth,clip=true,angle=0]{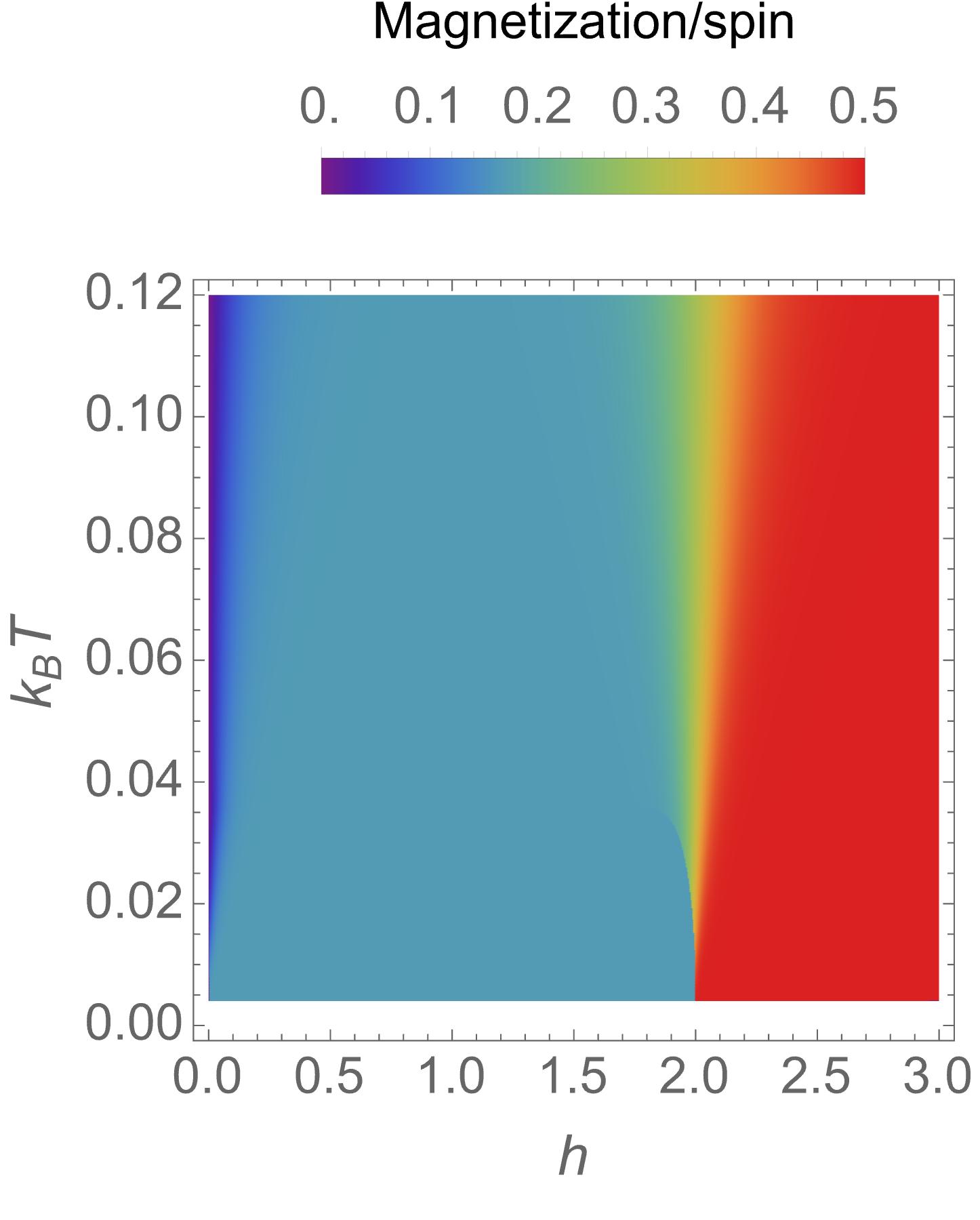}
        }
        \subfigure[][]{
\includegraphics[width=0.48\columnwidth,clip=true,angle=0]{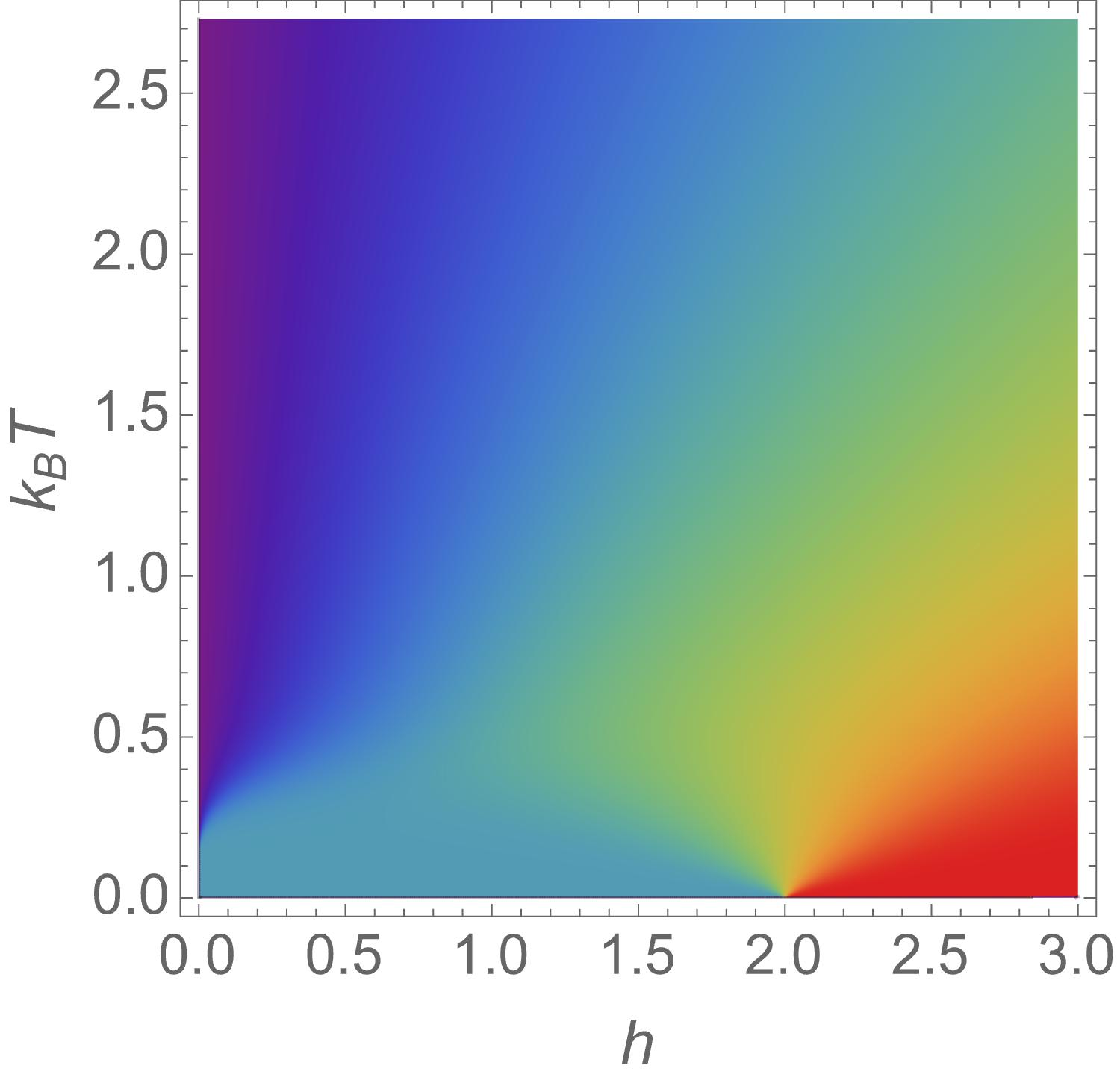}
        }
        \subfigure[][]{
\includegraphics[width=0.48\columnwidth,clip=true,angle=0]{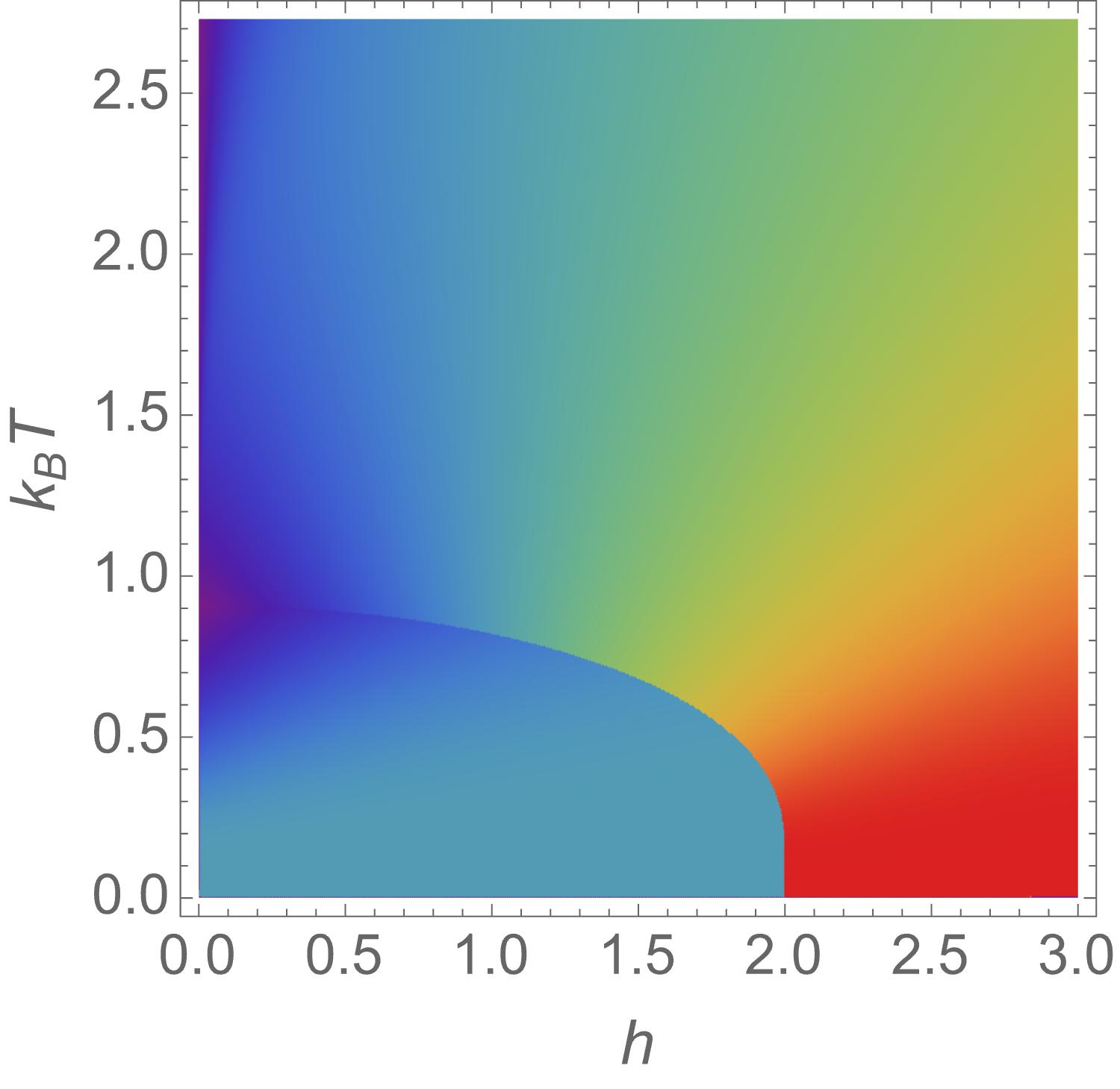}
        }
    \end{center}
\caption{\red{\figtextbf{The $T - h$ phase diagrams from the density plot of the magnetization per spin} $m/3=(\mu_a \langle \sigma_i \rangle + \mu_b \langle s_{i,1}+s_{i,2} \rangle)/3$ for (a) $J_2/J_1=1.95$ and $J=0$, (b) $J_2/J_1=1.95$ and $J=0.4S^2$, (c) $J_2/J_1=0$ and $J=0$, (d) $J_2/J_1=0$ and $J=20S^2$. $J_1=-S^2$ and $\mu_a=\mu_b=S$ where $S=1/2$.}}
\label{Fig:T-h-m}
\end{figure}


\vspace{0.5cm}
\noindent\textbf{Author Information} The author declares no competing interests.




\end{document}